\documentclass[english,aip, numerical,jcp,reprint,twocolumn,floatfix,prd]{revtex4-1}
\usepackage[flushleft]{threeparttable}
\usepackage{hyperref} 

\usepackage{graphicx}
\newcommand{\RRef}[1]{Ref.~\onlinecite{#1}}
\newcommand{\hof}{$\Delta H_f^\circ$}
\newcommand{\RRefs}[1]{Refs.~\onlinecite{#1}}

\newcommand{\INBLUE}[1]{\textcolor{black}{#1}}

\usepackage{graphicx}% Include figure files
\usepackage{dcolumn}% Align table columns on decimal point
\usepackage{bm}% bold math
\usepackage{amsmath}
\usepackage{multirow}
\usepackage{rotating}
\usepackage{mathpazo}
\usepackage{enumerate}

\usepackage{xcolor}

\begin{document}
	
\preprint{AIP/123-QED}
	
\title{
Critical Benchmarking of Popular Composite Thermochemistry
Models and Density Functional Approximations
on a Probabilistically 
Pruned Benchmark Dataset of Formation Enthalpies
} 

\author{Sambit Kumar Das}
\affiliation{Tata Institute of Fundamental Research, Centre for Interdisciplinary Sciences, Hyderabad 500107, India} 

\author{Sabyasachi Chakraborty}
\affiliation{Tata Institute of Fundamental Research, Centre for Interdisciplinary Sciences, Hyderabad 500107, India} 

\author{Raghunathan Ramakrishnan}
\email{ramakrishnan@tifrh.res.in}
\affiliation{Tata Institute of Fundamental Research, Centre for Interdisciplinary Sciences, Hyderabad 500107, India}


\date{\today}% It is always \today, today,

\begin{abstract}
\noindent 
First-principles calculation of the standard formation enthalpy, 
$\Delta H_f^\circ$~(298~K), in such large scale as required by chemical space explorations, is amenable only with density functional approximations (DFAs) and certain composite wave function theories (cWFTs). Alas, the accuracies of popular range-separated hybrid, `rung-4' DFAs, and cWFTs that offer the best accuracy-vs.-cost trade-off have until now been established only for datasets predominantly comprising small molecules; their transferability to larger systems remains vague. In this study, we present an extended benchmark dataset 
of $\Delta H_f^\circ$ for structurally and electronically diverse molecules. We apply quartile-ranking based on boundary-corrected kernel density estimation to filter outliers and arrive at Probabilistically Pruned Enthalpies of 
1,694 compounds 
(PPE1694). For this dataset, we rank the prediction accuracies of G4, G4(MP2), ccCA, CBS-QB3 and 23 popular DFAs using conventional and probabilistic error metrics. We discuss systematic 
prediction errors and highlight the role an 
empirical higher-level correction (HLC) 
plays in the G4(MP2) model. Furthermore, we comment on uncertainties associated with the reference empirical data for atoms and the systematic errors stemming from these
that grow with the molecular size. We believe these findings to aid in identifying meaningful application domains for quantum thermochemical methods. \end{abstract}

\maketitle

\section{\label{sec:levelintro}Introduction}
An unmet promise in ab-initio quantum chemistry is to predict molecular and reaction enthalpies with such high an accuracy, that reaction energies and relative stabilities of constitutional/conformational/geometric isomers can be established in agreement with experimentally observed trends---but at a computational cost that is comparable to that of a density functional approximation (DFA) with a reasonably converged basis set\cite{cioslowski2002quantum,irikura1998computational}. This situation is apparent in the context of emerging data science campaigns, where computational explorations and statistical inference of molecular properties across \emph{chemical compound space} is the prime focus\cite{von2018quantum,ramakrishnan2017machine,narayanan2019accurate,hachmann2011harvard,chakraborty2019chemical,arus2019exploring}. Arguably, one of the most sought-after molecular properties for data-mining is the standard formation enthalpy, $\Delta H_f^\circ$~(298~K), because of its significance to energetics and rates of industrial\cite{aston1942sources,duus1955thermochemical} and atmospheric chemical reactions\cite{rodgers1967estimation,dorofeeva2001nist}. Hence, development of a rapid thermochemistry protocol demonstrating a faithful transferability of accuracy---from that of the molecules used for the method's calibration ({\it i.e. training})---to a {\it new} molecule with arbitrary stoichiometry/valency and non-standard bond distances/angles will remain an active research domain\cite{curtiss2000reviews,curtiss2011gn,jensen2017introduction}.

Karton classified the most widely used composite wavefunction theories  (cWFTs) into those involving a post-CCSD(T)-level energy correction and those requiring a CCSD(T)-level treatment\cite{karton2016computational}. The former includes: W4\cite{karton2006w4}, HEAT-345QP\cite{bomble2006high}, and HEAT-456QP\cite{harding2008high} which depend 
on higher-order terms in the coupled-cluster expansion ({\it i.e.} quadruple, quintuple and higher excitations) to forecast molecular energies of `high chemical accuracy', with prediction error $\leq$ 1 kJ/mol)\cite{rogers2006heats,martin2005computational}. Alas, the severe computational complexities of these methods restrict their applicability to small molecules with at most a few dozen electrons. The latter category of thermochemistry methods is dependent on electron correlation energy estimated at the
relatively less intensive
CCSD(T)-level, where the triples contribution is included perturbatively. Such a compromise extends the application domain of this class of methods---for instance, CBS-QB3\cite{montgomery1994complete}, G4\cite{curtiss2007gaussian}, ccCA\cite{deyonker2006correlation,deyonker2006correlation_b,wilson2009correlation} and their offshoots---to larger molecules with up to a couple of dozen main group atoms. Depending on the size of the basis set with which the CCSD(T)
energy is estimated, cWFTs can yield an average prediction accuracy in the $\approx1-2$ kcal/mol range\cite{karton2016computational}. 
%using a finite basis set (as in CBS-QB3, G$n$, or ccCA) or through a complete-basis set (CBS) extrapolation (as in W1, W2\cite{martin1999towards}, W1U\cite{barnes2009unrestricted}) and other variations of Weizmann theories\cite{boese2004w3, karton2006w4, karton2012explicitlyWn, sylvetsky2016toward}, %cWFTs procedures  can yield an average prediction accuracy in the range $\approx1-2$ kcal/mol\cite{karton2016computational}. 
%More robust methods that are based on CCSDT(Q)/CBS such as W3-F12\cite{karton2012explicitlyWn}, W4-lite\cite{karton2006w4}, HEAT-345(Q)\cite{bomble2006high}, HEAT-456(Q)\cite{harding2008high} can provide an accuracy of $\approx$ 1 kJ/mol. Taking the accuracy even further, methods based on FCI/CBS such as by W4, W4.x, HEAT-345QP, and HEAT-456QP provide an accuracy of even less than 1 kJ/mol.
It must be noted that 
based on the computational complexities of these approaches, which
scale unfavorably with molecular size, 
their application domain is restricted to molecules of size 
CF$_4$ or CH$_3$COOH for FCI/CBS based methods, 
C$_6$H$_6$ or C$_6$H$_{14}$ for CCSDT(Q)/CBS based methods, 
C$_{20}$H$_{20}$ for CCSD(T)/CBS methods, while 
CCSD(T)/TZ based methods can handle large systems like C$_{60}$. 

A number of data-driven computational chemistry benchmark studies have explored subsets of the GDB17 molecular universe\cite{ruddigkeit2012enumeration} comprising 166,443,860,262 ({\it i.e.} 166.4 billion) closed-shell, organic molecules each containing up to 17 atoms of C, N, O, S, and halogens. These high-throughput computational studies have catered the data requirements of methodological studies focusing on machine learning (ML) modeling\cite{faber2018alchemical,lubbers2018hierarchical}. For instance, a past study computed DFT-level structures and properties of a small subset of the GDB17 dataset with 133,885 small organic molecules (the QM9 dataset), each containing up to nine C, O, N and F atoms\cite{ramakrishnan2014quantum,ramakrishnan2015big}. Recently, the QM9 dataset was subjected to  rigorous G4(MP2) treatment to estimate total energies, atomization energies and standard enthalpies of formation\cite{kim2019energy,narayanan2019accurate}. 
Other studies\cite{ward2019machine,dandu2020quantum} have extended these works by applying ML and 
$\Delta$-ML\cite{ramakrishnan2015big} to statistically infer G4(MP2)-level energies of a small set of organic molecules containing more than 9 heavy atoms. On the basis of dataset size alone, these studies represent some of the massive high-throughput quantum chemistry efforts ever undertaken. However, it remains to be seen if QM9 molecules' $\Delta H_f^\circ$ 
values predicted with G4(MP2) are quantitatively accurate 
to a degree that is relevant for comparison with experiments. 

The complexity involved in validating computed results in chemical space explorations is twofold: Firstly, it is a non-trivial task to automate the collection and pruning of  available experimental energies for such a large dataset as QM9. Secondly, even if all these molecules are `synthetically feasible', only a tiny fraction have been 
plausibly characterized by gas phase measurements. 
Seemingly, the only viable way of probing the transferability of a computational method to unexplored regions in the chemical space is to benchmark on compounds of similar chemical composition. 
In this context, Narayanan \textit{et al.}\cite{narayanan2019accurate} selected experimental 
values of $\Delta H_f^\circ$ for 459 closed-shell hydrocarbons 
and their substituted analogues---containing atoms similar to those in QM9---from
Pedley's extensive compilations\cite{pedley1986thermochemical,pedley1994thermochemical} 
and observed an average prediction error of 0.8 kcal/mol for G4(MP2). 
This value is comparable to that of G4(MP2)'s\cite{curtiss2007gaussian_a}
and ccCA's\cite{deyonker2009towards} mean errors noted for similar molecules in the G3/05 small molecules dataset\cite{curtiss2005assessment}. However, such high prediction accuracies are not expected to hold for electronically and structurally more diverse molecules that can be combinatorially generated from the QM9 set by protonation, deprotonation, or iso-valence-electronic substitutions. For instance, Schwilk \textit{et al.}. \cite{schwilk2020large} have selected about 4,000 molecules from the QM9 set and derived from this subset a new dataset QMspin, comprising 8000 triplet and 5000 singlet carbene compounds. 
To gain insight on the applicability of G4(MP2) to such non-trivial
chemical subspaces, it is a timely pursuit to benchmark widely employed cWFTs on larger, curated benchmark datasets by extending upon existing thermochemistry benchmark sets.

In this study we aim to: (i) collect reference $\Delta H_f^\circ$ values from several previous reports, and include new benchmark datasets containing experimental results and present a consolidated dataset, (ii) apply a probabilistic procedure based on the best theoretical method applicable to the entire set to detect and eliminate potential outliers, (iii) report on the prediction errors based on mean and percentiles metrics for the cWFTs, G4, G4(MP2), ccCA, and CBS-QB3 along with 23 popular DFAs and 2 semi-empirical methods, 
(iv) comment on the transferability of empirical corrections in G4(MP2) to the pruned 
$\Delta H_f^\circ$ dataset presented here,
%analyze the critical role of empirical corrections in G4(MP2) \INBLUE{"Now we just have discussed the transferability nothing on analysis of HLC"} and comment on the method's transferability, 
(v) from the larger dataset presented, identify and study a new benchmark set with isomerization reaction enthalpies, and (vi) finally, inspect the uncertainties in thermochemistry calculations arising due to the use of empirical reference data for atoms.
A general theme of our analyses is to shed more light on the transferability of
the G4(MP2) method that offers a suitable cost-accuracy trade-off for sampling across small molecules chemical spaces such as GDB17\cite{ruddigkeit2012enumeration}.

\section{Computational Details}
Thermochemistry calculations with G4(MP2), ccCA and DFAs
 were automated through \textit{in-house} scripts which
 rely on ORCA\cite{neese2012orca,neese2018software}. 
 Of the many ccCA variants developed by Wilson \textit{et al.}\cite{deyonker2006correlation,deyonker2006correlation_b,deyonker2009towards,deyonker2009accurate,laury2011pseudopotential,peterson2016prediction}, 
this work explores ccCA-P\cite{deyonker2006correlation_b}.
 G4 and CBS-QB3 calculations were carried out with Gaussian-16 suite of programs\cite{g16}.
PM6\cite{stewart2007optimization} and PM7\cite{stewart2013optimization} calculations were done via MOPAC\cite{MOPAC}.
We considered 23 DFAs from various levels of 
Perdew's `Jacob's ladder'\cite{perdew2001jacob}:  
generalized gradient approximation (GGA)---BLYP\cite{becke1988density}, PW91\cite{perdew1992k} and PBE\cite{perdew1996phys};
hybrid GGA---B3LYP\cite{becke1993becke}, O3LYP\cite{cohen2001dynamic}, X3LYP\cite{xu2004x3lyp} and PBE0\cite{adamo1999toward}; 
meta-GGA---TPSS\cite{tao2003climbing};
hybrid meta-GGA---TPSS0\cite{grimme2005accurate} and M06-2X\cite{zhao2008m06}. 
We also selected the range-separated hybrid functionals:
CAM-B3LYP\cite{yanai2004new}, $\omega$B97X\cite{chai2008long}, $\omega$B97X-D3\cite{lin2013long}, $\omega$B97X-V\cite{mardirossian2014omegab97x}, $\omega$B97M-V\cite{mardirossian2016omega}, $\omega$B97X-D3BJ, and $\omega$B97M-D3BJ, 
where Grimme's D3 dispersion with Becke-Johnson damping (D3BJ)\cite{grimme2011effect,becke2005density,johnson2005post,johnson2006post} has been included explicitly. 
Furthermore, B2PLYP\cite{grimme2006semiempirical}, B2PLYP-D3\cite{grimme2010consistent} and mPW2PLYP-D\cite{schwabe2007double} represent the double-hybrid functionals considered in this study, some of which include dispersion corrections. Additionally, we also
benchmarked a few of the aforestated functionals with an additional dispersion correction, namely, B3LYP-D3, TPSS0-D3 and M06-2X-D3.

Initial geometries of molecules from previously studied datasets were collected from their corresponding sources---when such information was available---and subjected to geometry relaxations. For compounds with no previously reported geometries, we consulted popular online chemical databases: NIST\cite{NIST}, ChemSpider\cite{pence2010chemspider} and Pub-Chem\cite{kim2019pubchem}. In some cases, we created initial geometries using the software  Avogadro\cite{hanwell2012avogadro} and carried out minimum energy geometry relaxation with universal forcefield (UFF)\cite{rappe1992uff}. We performed DFA calculations at the B3LYP/6-31G(\textit{2df,p}) minimum energy geometry in a single point fashion using the def2-QZVP basis set, which has been shown to yield predictions close to the Kohn-Sham limit\cite{weigend2005balanced}. PM6 and PM7 calculations were performed using {\tt precise} geometry relaxation thresholds at the corresponding levels. Geometry optimizations with DFAs were carried out with {\tt tight} convergence criteria with $10^{-4}$ as the threshold for the maximum component of the force vector. To facilitate convergence of the geometry towards the energy minimum, force constants were computed at every fifth step of geometry optimization. In all calculations, SCF convergence was attained with {\tt verytightscf} criteria corresponding to a threshold of $10^{-9}$ Hartree for the total energy. For the numerical quadrature of the exchange correlation part of the energy in DFAs, we used Lebedev-434 angular grids and {\tt Grid7} settings for Gauss-Chebyshev radial grids. From the zero-point corrected electronic energy, $\Delta H_f^\circ$~(298K) was calculated according to the standard convention\cite{ochterski2000thermochemistry}. Heats of formation of atoms at $0$ K, $\Delta H_f^\circ (0$ K$)$, and enthalpy corrections, for elements in their standard states, $H^\circ(298$ K$) - H^\circ(0$ K$)$ are listed in APPENDIX. Spin-orbit corrections to the electronic energies of atoms, ions, selected diatomic molecules\footnote{Note that in Table I of \RRef{curtiss2001extension} SeH$^+$ has to be SeH.} and acetylene were collected from \RRefs{curtiss1995extension,curtiss2001extension,curtiss2007gaussian}.

\begin{figure*}[htbp]
    \centering
    \includegraphics[width=17.8cm]{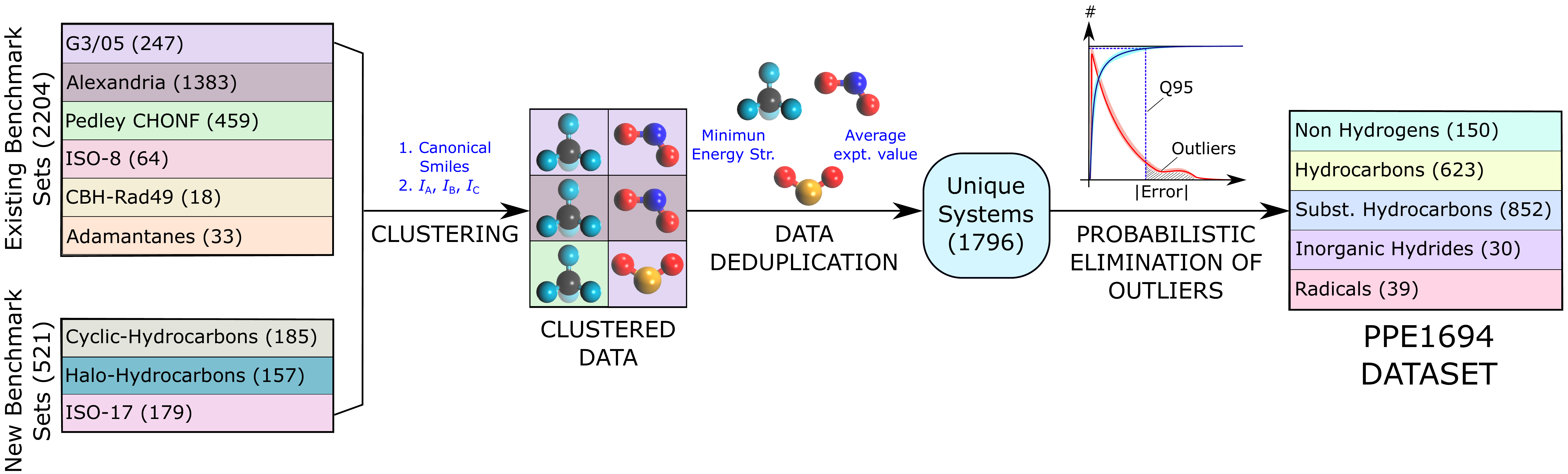}
    \caption{PPE1694 dataset of molecular standard formation enthalpies: 
    Data collection, deduplication and probabilistic filtering of
    outliers. See text for more details of the individual datasets and their sources.}
    \label{fig:ppe}
\end{figure*}

\section{Benchmark Dataset of molecular standard formation enthalpies}\label{sec:dataset}
%\subsection{Data collection}
We gathered benchmark values of $\Delta H_f^\circ$ from six previously studied datasets amounting to 2,204 entries: 
\begin{enumerate}
\item G3/05 dataset\cite{curtiss2005assessment} is a set of 454 energies distributed into: 270 \hof~, 105 ionization energies (IE), 63 electron affinities (EA), 10 proton affinities (PA), and 6 binding energies (BE) of hydrogen-bonded complexes. This set has been extensively used in the development of the G$n$ theories and also for validating other \textit{ab initio} thermochemical protocols. We note in passing that in the 270 subset only 247 are \hof{} values, the remaining 23 are atomization energies that we excluded. 

%\INBLUE{that were not considered here}. 

\item Alexandria dataset\cite{ghahremanpour2018alexandria} contains classes of molecules similar to that of G3/05 but many in number with diverse molecular sizes, thus, providing a platform to examine the transferability of methods that perform well for G3/05. Out of 2,704 entries in this dataset, 1,383 compounds that contain experimental values of \hof~ have been selected. % 1392 --> 1383

%\INBLUE{We selected 1392 entries with experimental values, "shall we put a comment saying a sub-set of Alexandria was validated with G4?"}. 

\item Pedley CHONF is a set derived from 459 experimental values for hydrocarbons and substituted hydrocarbons from Pedley's report\cite{pedley1994thermochemical} containing the atoms H, C, N, O, and F. An earlier study\cite{narayanan2019accurate} benchmarked the accuracy of G4(MP2) for this dataset and reported an average prediction error of 0.8 kcal/mol.  
%For 4 stoichiometries, we have considered both {\it cis} and {\it trans} isomers taking the tally to 463.
%\INBLUE{"different no. of systems (459 and 463 respectively), Is a comment required for that?"}

\item ISO-8 comprises experimental isomerization energies and the corresponding $\Delta H_f^{\circ}$ 
for 8 sets of constitutional isomers amounting to 64 entries\cite{schreiner2006many}.

\item CBH-Rad49\cite{sengupta2014prediction} contains 49 cyclic \& acyclic radicals with reference $\Delta H_f^{\circ}$ energies gathered from experiment and high-level theoretical modeling. From this set, we collect 18 experimentally determined entries. 

%CBH-rad49\cite{sengupta2014prediction} contains a collection of 49 cyclic \& acyclic radicals with diverse functional groups and hybridizations. Reference $\Delta H_f^{\circ}$ energies in this dataset are collected from a wide range of resources, combining values from experiments and high-level theoretical modeling. \INBLUE{From the 49 entries only 18 entries were considered in this study, the remaining 31 radical systems were neglected, since the reference \hof~ values were taken from theoretical evaluations}

\item $\Delta H_f^{\circ}$ of 33 adamantanes estimated by a simultaneous least-squares regression involving a thermochemical network of 300 isodesmic reactions\cite{dorofeeva2018self}.

%\item Dorofeeva \textit{et al.}\cite{dorofeeva2018self} have presented a dataset of 33 adamantanes for which $\Delta H_f^{\circ}$ were estimated by simultaneous least-squares regression of a thermochemical network containing 300 isodesmic reactions.

\end{enumerate}

%Among all the datasets considered here, only G3/05 has been previously studied with DFT, WFTs and cWFTs\cite{curtiss2007gaussian_a,deyonker2009towards,curtiss2005assessment}, hence, it serves as a critical reference to evaluate the accuracy of thermochemistry methods. 

To further enrich the benchmark set collected from previous findings, we included 521 compounds from \RRefs{pedley1986thermochemical,pedley1994thermochemical} that have
never been subjected to theoretical modeling.These compounds belong to the following three categories: cyclic hydrocarbons, halogen-rich hydrocarbons and constitutional isomers with 17 unique stoichiometries (ISO-17). The number of compounds added from each of the aforestated dataset is shown in Fig.~\ref{fig:ppe}. 
The total number of compounds from the collective set amounts to 2,725, where we detected a number of repeated entries. To eliminate redundant entries, we followed a systematic data-deduplication procedure based on stoichiometries, principal moments of inertia and counts of heavy atom bonds. Firstly, we separated radicals from non-radicals, and binned entries in each category 
appropriately based on stoichiometries. Following this, we clustered constitutional isomers for a given stoichiometry based on canonical smiles and principal moments of inertia of every molecule; each cluster is considered as a set of redundant entries. Finally, for each cluster, 
the average experimental value was calculated and the conformer with the least G4(MP2) value;
for ambiguous cases, we used G4 results.

%For a detailed description of the dataset kindly consult the SI.
%When multiple values of \hof~ differing by $>$ 0.1 kcal/mol were found for the same compound, we have retained the more frequently studied entry. \INGREEN{To eliminate the redundant systems, we followed a data deduplication procedure, where the first step was to cluster these redundant systems using canonical smiles and principal moments of inertia. For each cluster the average experimental value was considered and the structure corresponding to the lowest calculated(G4(MP2)) value was taken into consideration.}
%2796\INBLUE{(2,725)}
%To eliminate the redundant systems, we followed a data deduplication procedure, where the first step was to cluster these redundant systems using canonical smiles and principal moments of inertia. For each cluster the average experimental value was considered and the structure corresponding to the lowest calculated(G4(MP2)) value was taken into consideration.

Following de-duplication, 2,725 entries reduced to 1,796 unique entries, for which
%For the resulting dataset of 2021\INBLUE{(1,796)} compounds,
it is a non-trivial task to quantify---{\it ab initio}---plausible uncertainties arising from random and systematic errors in the experimental measurements. Paulechka \textit{et al.}\cite{paulechka2017efficient} discussed the reasons for typical uncertainties in experimentally determined \hof{} and indicated that such deviations can amount to even a few kJ/mol. 
In that study, for very critical benchmarking of DLPNO-CCSD(T) \hof, the authors selected 45 compounds with at least two independent experimental results. 
Even when using such `precise' experimental values as references, error trends based on the mean unsigned error (MUE) can severely underestimate thermochemical uncertainty\cite{ruscic2014uncertainty}. 
Simm \textit{et al.}~\cite{simm2017error} have discussed how a performance analysis based on MUE is prone to fail as it does not distinguish the systematic contributions to the errors from the non-systematic counterparts. For pathological error distributions, arising plausibly due to the presence of outliers in the reference dataset, prediction uncertainties can be truncated using percentile-based error metrics \cite{ruscic2014uncertainty,pernot2018probabilistic,thakkar2015well,wu2015choosing}.
More specifically, performance ranking of \textit{ab initio} methods revealing trends in uncertainties
has been made possible by the use of 
the 95$^{\rm th}$ percentile of the absolute intensive ({\it i.e.} normalized)
error distributions along with the
MUE\cite{pernot2020probabilistic}. It is the subject of 
Section~\ref{subsec:prob} to 
discuss a procedure to detect probabilistic tendencies of an entry
from the 1,796 set to be an outlier.
%2021\INBLUE{(1,796)}
%\INBLUE{Following systems have structural issues in them
%\begin{itemize}
%    %\item G3-05/05-EOF-Radicals/17-CH3CH2O (Imaginary Freq.)
%    \item G3-05/07-IP-molecule/IP-mol-00025 (Imaginary Freq. -15)
%    \item G3-05/09-EA-molecule/EA-mol-00043 (Imaginary Freq. -240 neutral)
%    \item Alexandria/0322-26-dimethylhept-3-yne (Optimization-not converged, small imaginary)
%    %\item Alexandria/1096-methylcyclopentane (Optimization-not converged)
%    %\item ISO-8/ISO-C10H10/008-4-Cyclopropylbuta-1-3-diyn-1-yl-cyclopropane (Optimization-not converged)
%    \item ISO-8/ISO-C6H6/004-Hexa-2-4-diyne (ran out of cycle )
%\end{itemize}
%}

\section{Probabilistic Pruning of the Dataset}\label{subsec:prob}
For the duplicate-free 1,796 set, we utilize a probabilistic argument to 
detect outliers. %2021\INBLUE{(1,796)}
The principle behind this scheme
is to use highly accurate computed values of \hof~  as references and 
mark only those molecules that lie beyond the 95$^{\rm th}$ percentile of the
error distribution as outliers. The cumulative probability used for this purpose acts as a 
{\it prior} distribution---more robust the reference, more reliable
the prior is. Hence, it may be anticipated that this scheme is 
guaranteed to detect genuine outliers when one of the high-precision methods 
such as W4, HEAT-456QP, or HEAT-345QP 
is used as a reference. Given the  
size of majority of the molecules in the dataset, we depend on G4 
values to define the prior probability. %G4(MP2)\INBLUE{(G4)}
For 20 electron-rich systems in the 1,796 set, for which 
G4 calculations were not amenable, we relied on G4(MP2). 
%Further computational details of the calculations performed are deferred to Section~\ref{sec:comp} 
%\INBLUE{Unfortunately due to the high computation-cost, G4 energies could not be obtained for 20 systems, for which corresponding G4(MP2) energies were used}; computational details of the calculations are deferred to Section~\ref{sec:comp}.

\begin{figure}[htbp]
    \centering
    \includegraphics[width=8.6cm]{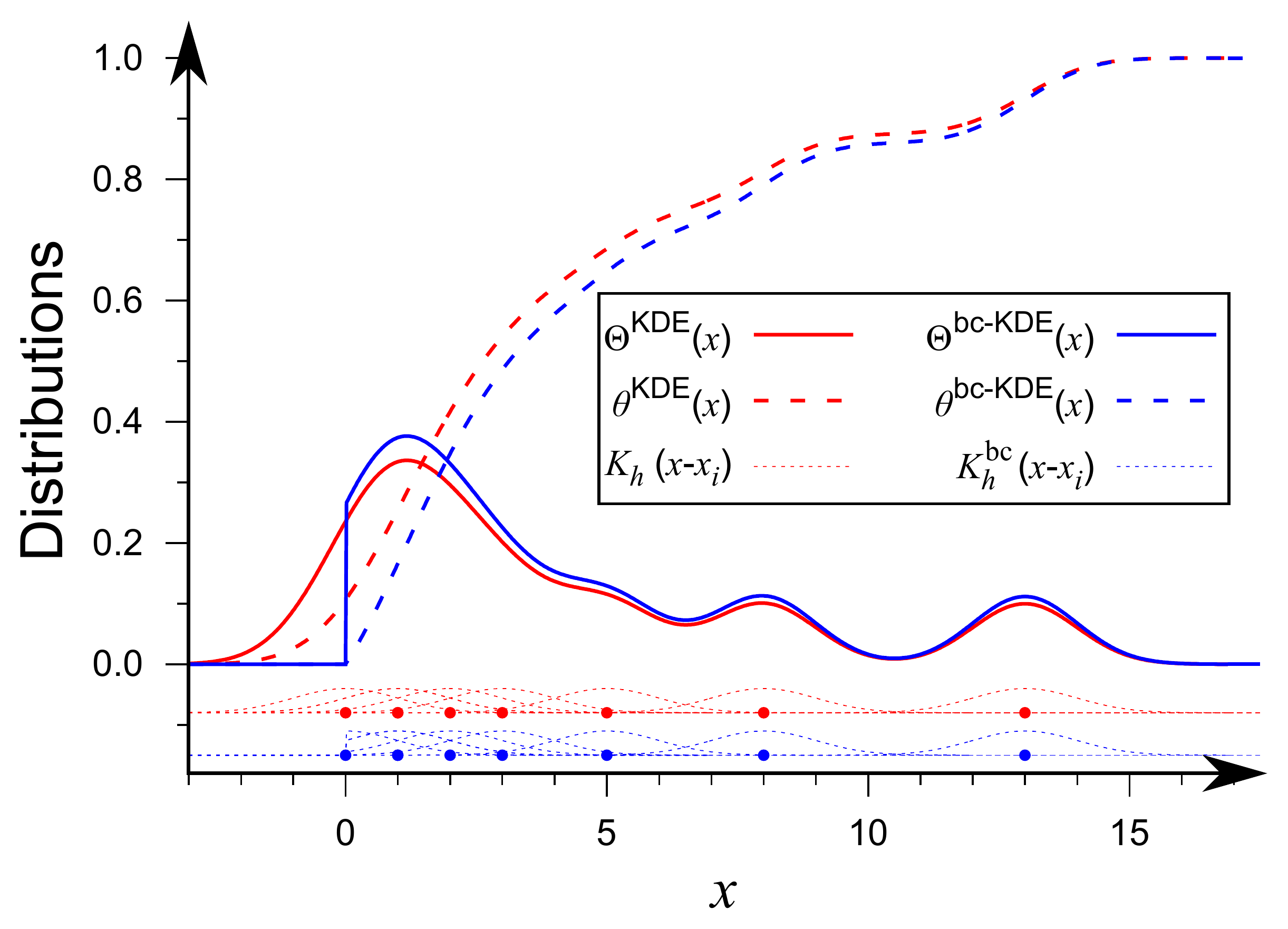}
    \caption{Probability density function, $\theta(x)$, and
cumulative distribution function, $\Theta(x)$, based on kernel
density estimation (KDE) and its boundary corrected analog, bc-KDE,
demonstrated for an exemplary dataset with eight points:
$x_1=0$, $x_2=x_3=1$, $x_4=2$, $x_5=3$, $x_6=5$, $x_7=8$ and $x_8=13$.
For clarity, $\theta(x)$ has been multiplied by 100.
Also shown in the bottom of the plot are the kernel basis functions
$K_h(x-x_i)$ and $K_h^{\rm bc}(x-x_i)$ centered at the data points
with arbitrarily shifted ordinates.}
    \label{fig:App1}
\end{figure}

We begin with absolute errors in 
G4-predicted $\Delta H_f^\circ$ for the 1,796 set. %G4(MP2)\INBLUE{(G4)} 2021\INBLUE{(1,796)}
Following the arguments presented by Savin \textit{et al.}\cite{savin2014judging} and Perdew \textit{et al.}\cite{perdew2016intensive}, and selected an intensive error measure 
to account for the fact that the dataset contains molecules spanning various sizes. For this purpose, we used MUE per valence electron, to capture periodic trends in molecular thermochemistry. From discrete values of the G4 error, we obtained a continuous probability %G4(MP2)\INBLUE{(G4)}
distribution using the boundary corrected kernel density estimation (bc-KDE), where a radial basis function is expanded at each discrete value. 
In KDE, we take a kernel in the form of the standard normal distribution
${\mathcal N}(0,1)$
\begin{eqnarray}
  K_h \left( x-x_i \right) = \frac{1}{\sqrt{2 \pi} h} \exp \left[ - \frac{\left( x - x_i\right)^2}{2 h^2} \right],
\end{eqnarray}
where $K_h$ is the normalized estimator. 
A kernel function is expanded at every data point, giving
rise to the (unnormalized) total density function that 
is an average of all kernels
\begin{eqnarray}
  \hat{f}(x,h) & = & \frac{1}{N} \sum_{i=1}^N K_h \left( x-x_i \right)
\end{eqnarray}
It is a well-known problem that KDE can delocalize beyond the allowed
domain; in such cases, the naive approach of truncating the total density function 
$f(x)$ often underestimates the actual probability distribution.
This effect is illustrated using a toy dataset 
in Fig.~\ref{fig:App1}.  
To this end, we employ boundary-corrected kernels of the form
\begin{eqnarray}
K_h^{\rm bc} \left( x-x_i \right) & = & \left [ K_h \left( x-x_i \right) + K_h \left( x+x_i-2x^* \right) \right] \mathbf{1}_{x \in A}, \nonumber \\
& &
\end{eqnarray}
where the indicator function $\mathbf{1}_{x \in A}$ is 1 when $x \in A$ and vanishes otherwise\cite{gramacki2018nonparametric}.
Bc-KDE captures the true nature of the probability density when the property is bounded. While the individual kernel functions that cross the boundary are truncated by the indicator function, the total probability is still conserved through a normalization of the resulting probability density, see Eq.~\ref{eq:normpdf}.

In our bc-KDE calculations, we used a kernel width of 0.005 kcal/mol and determined the total density function as the average
\begin{eqnarray}
  \hat{f}(x,h) & = & \frac{1}{N} \sum_{i=1}^N \left [ K_h \left( x-x_i \right)
+ K_h \left( x+x_i-2x^* \right) \right] \mathbf{1}_{x \in A}, \nonumber\\
& &
\end{eqnarray}
which is normalized over the domain to give the
probability density function (PDF)

\begin{eqnarray}
  \hat{\phi}(x) & = & \frac{ \hat{f}(x,h) }{ ||\hat{f}(x,h)||_2 }.
  \label{eq:normpdf}
\end{eqnarray}
The cumulative density function (CDF) used to identify percentiles 
is obtained by integrating the PDF
\begin{eqnarray}
  \hat{\Phi}(x)  = \int_{A}^{B} dx\,\hat{\phi}(x).
\end{eqnarray}

\begin{figure}[htbp]
    \centering
    \includegraphics[width=8.6cm]{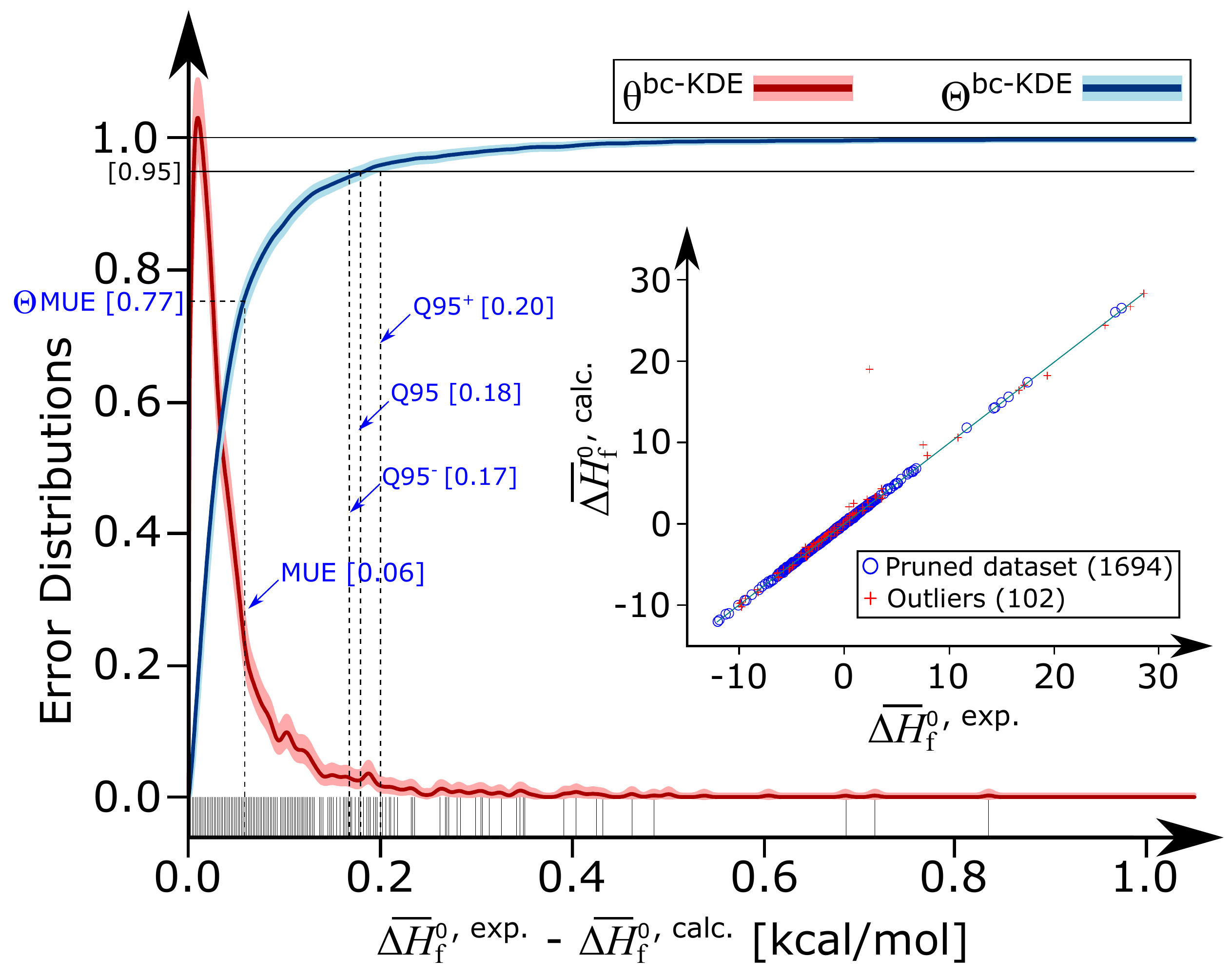}
    \caption{Boundary-corrected kernel density probability distribution,
    $\theta^{\mathrm{bc-KDE}}$ and the corresponding cumulative density function $\Theta^{\mathrm{bc-KDE}}$ for the absolute deviations of the calculated \hof{}.
    Calculated results comprise G4 values for 1,674 small-to-medium systems, and 
    G4(MP2) values for 20 electron-rich molecules.
    The overline indicates that%2021\INBLUE{(1,796)}
    the values are normalized over valence electrons. Uncertainty envelopes were determined by bootstrapping.
    The inset features a scatterplot. 
    }
    \label{fig:CDF}
\end{figure}

The resulting CDF enables statistical modeling of the probability to obtain a certain degree of prediction accuracy for a given quantum chemistry method as previously noted by Pernot {\it et al.}\cite{pernot2018probabilistic}. To estimate the variance of the CDF, we followed bootstrapping with
1000 shuffles, with 500 points randomly sampled from the 1,796 set. The boundary-corrected kernel density probability distribution, $\theta^{\mathrm{bc-KDE}}$ and the corresponding cumulative density function $\Theta^{\mathrm{bc-KDE}}$ are on display in Fig.~\ref{fig:CDF}. To eliminate any sampling bias, we use the lower bound of the 95$^{\rm th}$ percentile, denoted $Q95^-$ in Fig.~\ref{fig:CDF}, as a threshold; all compounds for which the MUE per valence electron is greater than $Q95^-$ are marked as outliers and eliminated from the dataset. We thus arrive at the benchmark set of probabilistically pruned enthalpies of 1,694 compounds, denoted henceforth PPE1694. 
%The corresponding MUE of the G4 method  1.94 kcal/mol, dropping to 1.47 kcal/mol for the pruned 1,674 set.
For the original 1,796 set, MUE of G4(MP2) is 2.17 kcal/mol, for 1,776 entries G4's 
MUE is 1.94 kcal/mol. For the pruned 1,694 set, G4(MP2)'s MUE is 1.70 kcal/mol. As stated
before, G4 calculations were not amenable to 20 electron-rich systems, hence for the collective
reference set with 1,674 G4 and 20 G4(MP2) values, the MUE is 1.51 kcal/mol.

%In the pruned set, we have 1,674 G4 entries and 20 G4(MP2) ones.

%In this way, elimination of 102 outliers from the original 1,796 set, decreased G4(MP2)'sMUE 
%\INBLUE{Incomplete: G4(MP2):2.17 to 1.70. G4:1.94(for 1,776 systems) to 1.47(for 1,674). Pruned data(1,674G4 and 20 G4(MP2)): 1.98 to 1.51)} % PPE1908\INBLUE{(PPE1694)} 113\INBLUE{(102)} 2021\INBLUE{(1694)}
%MUE for the dataset from 2.17 kcal/mol to 1.70 kcal/mol. %2.10\INBLUE{(2.17)} 1.65\INBLUE{(1.70)}
%This final MUE is 0.71 kcal/mol larger %0.66\INBLUE{(0.71)}
%than the method's MUE for G3/05's 270 subset. 
%The corresponding MUE of the G4 method 
%1.94 kcal/mol, dropping to 1.47 kcal/mol for the pruned 1,674 set.

\begin{figure}[htbp]
    \centering
    \includegraphics[width=8.6cm]{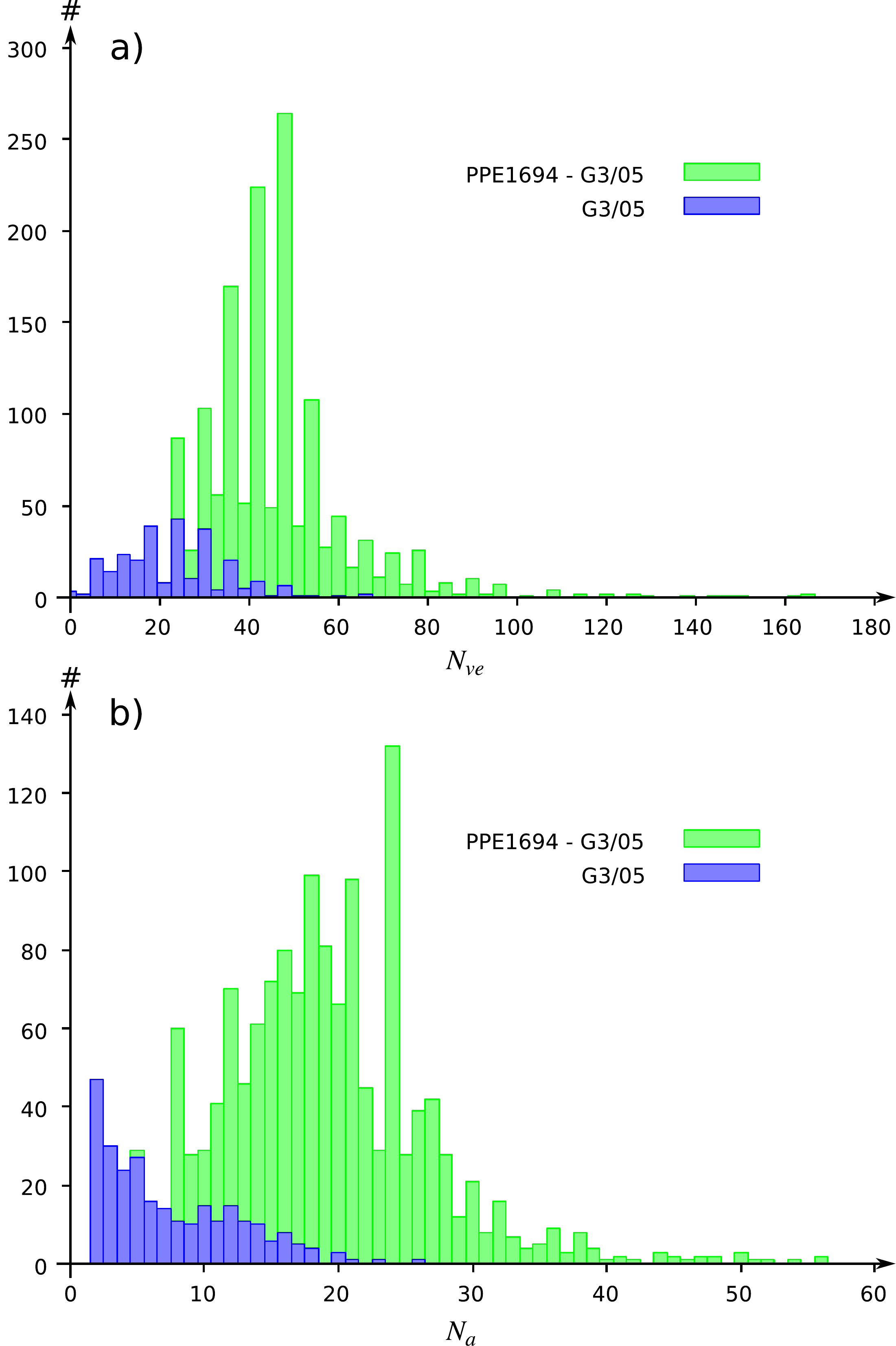}
    \caption{Distribution of compounds in the G3/05 dataset compared to other molecules that are present in PPE1694: a) Counts for number of valence electrons%PPE1908\INBLUE{(PPBE1694)}
    $N_{ve}$, b) Counts for number of atoms $N_a$.}
    \label{fig:datasets}
\end{figure}

In Section\ref{sec:sys}, we discuss how the performances
of cWFTs and DFAs deteriorate systematically 
because of the uncertainties associated with
empirical atomic parameters used for thermochemistry energetics.
Compared to the small molecules benchmark set G3/05, PPE1694 covers the space of molecules that contain more atoms as well as more valence electrons; %\INBLUE{(PPBE1694)}
qualitative trends are illustrated in Fig.~\ref{fig:datasets}. The largest number of valence electrons in G3/05 and PPE1694 sets are 66 (C$_6$F$_6$ and %PPE1908\INBLUE{(PPBE1694)}
C$_6$F$_5$Cl) and 166 (C$_{10}$F$_{18}$),  respectively. On the other hand, the maximum number of atoms in a given compound is 26 (C$_8$H$_{18}$) for the G3/05 set, and 56 (C$_{18}$H$_{37}$Cl) for PPE1694. Both sets are dominated by closed-shell %PPE1908\INBLUE{(PPBE1694)}
cases as evident from larger counts for compounds with even number of valence electrons, see Fig.~\ref{fig:datasets}a.

\section{Dressed atom corrections for systematic errors in density functional approximations}
\INBLUE{It is well known that DFA predicted \hof{} suffer systematic errors through insufficient modeling of atomization energies. Such errors can be empirically corrected either through a quasi-atom correction scheme\cite{winget2004B3LYP,grimme2005accurate} or by a bond density function approach\cite{cioslowski2000aset}. The latter method captures the atomic local environment in terms of chemical bonds yielding a parameterization guided by the unique chemistry of the molecule. However, this often results in overfitting to a given dataset. Here, we rely on an additive quasi-atom
correction scheme because of its inherent robustness and simplicity. Accordingly, 
DFA predicted \hof{} is corrected by a sum of element-specific constant
\begin{equation}
 \Delta H_f^\circ(\mathrm{exp.}) = \Delta H_f^\circ(\mathrm{DFA}) + \sum_i c_i n_i,    
\end{equation}
where $c_i$ is the correction for the $i$-th atom-type while $n_i$ is the number of such atoms in the molecule. From PPE1694, we randomly selected 300 entries ensuring all elements were represented at least twice. This set was hence used to determine the element-wise coefficients  
through a least-squares regression.  
%In this study, we parameterize all 23 functionals and compile the results along with composite protocols and semi-empirical methods in Table.~\ref{table:big}. 
Coefficients for all 22 unique elements present in PPE1694 are available, for 23 DFAs studied here, on the Supplementary Information page \href{http://moldis.tifrh.res.in/data/prunedHOF}{http://moldis.tifrh.res.in/data/prunedHOF}.}

\begin{turnpage}
\begin{table*}[h]
\centering
\caption{Prediction errors in the modeling of \hof. For selected cWFTs, DFAs and semi-empirical methods, mean unsigned error (MUE) is reported for each class of compounds in the PPE1694 benchmark dataset. The root-mean-square-error is given in parenthesis while, the most positive and negative errors are provided inside a square bracket. All values are in kcal/mol.}
\begin{threeparttable}
\resizebox{1.3\textwidth}{!}{
\begin{tabular}{ l r r r r r r }
    \hline
Methods  &             Nonhydrogens (150)   &         Hydrocarbons (623)   &     Subst. hydrocarbons (852) &  Inorganic hydrides (30) &  Radicals (39) &  Total (1,694) \\
\hline
& & & & & & \\
G4(MP2) & 2.21 (3.31) [-12.33, 11.15]  & 1.81 (2.80) [-10.47, 18.08]  & 1.57 (2.15) [-7.25, 9.81]   & 0.85 (1.20) [-4.15, 1.96]   & 1.37 (1.88) [-3.35, 5.31]  & 1.70 (2.50) [-12.33, 18.08] \\ 

G4\tnote{a} & 1.95 (2.97) [-11.05, 11.73]  & 1.51 (2.32) [-10.75, 11.02]  & 1.40 (1.98) [-7.06, 9.16]   & 0.98 (1.25) [-3.87, 1.65]   & 1.13 (1.49) [-3.45, 3.47]  & 1.47 (2.18) [-11.05, 11.73] \\

ccCA\tnote{b} & 2.15 (3.24) [-12.54, 10.50]  & 1.69 (2.43) [-9.60, 19.71]   & 1.58 (2.18) [-9.36, 8.67]   & 1.00 (1.27) [-2.52, 2.99]   & 1.59 (2.25) [-5.88, 1.03]  & 1.66 (2.37) [-12.54, 19.71] \\

CBS-QB3\tnote{c} & 6.29 (8.87) [-14.88, 31.23]  & 2.75 (3.41) [-15.99, 6.76]   & 2.32 (3.38) [-6.74, 17.25]  & 2.87 (4.41) [-17.14, 9.63]  & 1.70 (2.20) [-6.00, 2.82]  & 2.82 (4.17) [-17.14, 31.23] \\

\hline
& & & & & & \\

BLYP & 7.22 (11.37) [-40.86, 38.13] & 7.53 (11.63) [-74.83, 16.85] & 6.24 (9.93) [-77.27, 10.34] & 4.19 (6.19) [-24.98, 8.43]  & 6.35 (8.42) [-27.01, 10.91]& 6.77 (10.63) [-77.27, 38.13] \\

PW91 & 6.02 (8.53) [-25.20, 35.93]  & 3.06 (4.44) [-27.32, 9.77]   & 3.74 (5.23) [-29.15, 8.49]  & 4.35 (5.69) [-15.72, 10.56] & 3.58 (4.30) [-8.18, 9.51]  & 3.70 (5.33) [-29.15, 35.93] \\

PBE & 6.26 (8.80) [-25.67, 36.33]  & 3.07 (4.42) [-26.75, 9.77]   & 3.93 (5.47) [-29.14, 8.09]  & 4.53 (6.02) [-16.35, 11.30] & 3.70 (4.47) [-7.89, 9.89]  & 3.83 (5.49) [-29.14, 36.33] \\

TPSS & 6.18 (9.15) [-31.67, 34.61]  & 4.75 (7.15) [-44.02, 11.08]  & 5.01 (7.22) [-46.26, 7.44]  & 5.18 (6.93) [-19.76, 13.62] & 4.48 (5.54) [-13.10, 10.60]& 5.01 (7.35) [-46.26, 34.61] \\

\hline
& & & & & & \\

B3LYP & 4.67 (7.05) [-26.39, 18.00]  & 6.11 (9.16) [-55.34, 10.12]  & 3.68 (6.30) [-55.38, 7.48]  & 2.55 (3.57) [-12.54, 5.43]  & 4.75 (6.25) [-19.85, 8.29] & 4.67 (7.51) [-55.38, 18.00] \\

B3LYP-D3 & 3.71 (5.27) [-15.76, 17.90]  & 3.34 (4.95) [-26.15, 11.83]  & 2.45 (3.65) [-25.26, 9.37]  & 2.24 (3.10) [-10.08, 6.03]  & 3.49 (4.30) [-11.22, 6.47] & 2.91 (4.33) [-26.15, 17.90] \\

O3LYP & 4.81 (6.56) [-19.77, 21.05]  & 3.60 (4.88) [-28.19, 11.15]  & 3.00 (4.10) [-23.34, 11.24] & 3.46 (4.44) [-8.55, 8.02]   & 2.97 (3.98) [-10.82, 9.70] & 3.39 (4.66) [-28.19, 21.05] \\

X3LYP & 4.34 (6.34) [-22.66, 16.34]  & 5.70 (8.54) [-51.25, 9.49]   & 3.31 (5.70) [-50.92, 8.01]  & 2.45 (3.36) [-11.22, 5.37]  & 4.56 (5.92) [-18.44, 8.10] & 4.29 (6.91) [-51.25, 16.34] \\

PBE0 & 3.91 (5.22) [-12.84, 18.56]  & 3.17 (4.14) [-21.31, 10.39]  & 2.34 (3.17) [-13.38, 10.61] & 2.79 (3.76) [-8.08, 7.22]   & 3.11 (3.83) [-9.22, 10.47] & 2.81 (3.79) [-21.31, 18.56] \\

\hline
& & & & & & \\

TPSS0 & 4.00 (5.34) [-17.61, 17.57]  & 4.47 (5.96) [-30.93, 8.85]   & 2.86 (3.96) [-26.08, 9.71]  & 3.64 (4.36) [-7.34, 8.62]   & 4.15 (4.66) [-5.94, 10.30] & 3.60 (4.93) [-30.93, 17.57] \\

TPSS0-D3 & 3.53 (4.80) [-12.30, 18.25]  & 3.13 (3.91) [-14.30, 17.48]  & 2.24 (2.97) [-11.33, 12.05] & 3.68 (4.29) [-6.76, 9.21]   & 3.44 (4.01) [-6.77, 9.82]  & 2.73 (3.58) [-14.30, 18.25] \\

M06-2X & 2.99 (4.15) [-12.77, 14.39]  & 3.45 (4.31) [-21.81, 6.78]   & 2.52 (3.43) [-11.49, 16.40] & 1.51 (1.97) [-5.63, 3.93]   & 2.11 (2.86) [-7.83, 6.43]  & 2.88 (3.81) [-21.81, 16.40] \\

M06-2X-D3 & 3.01 (4.18) [-12.80, 14.39]  & 3.29 (4.08) [-19.70, 7.03]   & 2.51 (3.42) [-9.75, 16.57]  & 1.51 (1.97) [-5.62, 3.92]   & 2.11 (2.84) [-7.69, 6.35]  & 2.81 (3.72) [-19.70, 16.57] \\

\hline
& & & & & & \\

$\omega$B97X & 2.90 (3.97) [-10.54, 16.67]  & 2.38 (3.16) [-15.27, 11.30]  & 1.86 (2.47) [-7.84, 12.19]  & 2.13 (2.67) [-5.09, 6.38]   & 2.38 (3.10) [-7.33, 10.27] & 2.16 (2.91) [-15.27, 16.67] \\

$\omega$B97X-D3 & 3.00 (4.09) [-11.41, 17.44]  & 2.28 (2.97) [-13.06, 11.06]  & 1.86 (2.46) [-8.13, 11.17]  & 2.08 (2.60) [-5.70, 5.72]   & 2.31 (2.99) [-7.42, 9.70]  & 2.13 (2.85) [-13.06, 17.44] \\

$\omega$B97X-V & 3.28 (4.75) [-11.04, 20.83]  & 2.74 (3.49) [-12.27, 15.93]  & 2.11 (2.91) [-7.04, 14.40]  & 2.55 (3.20) [-5.72, 8.76]   & 2.44 (3.43) [-7.98, 12.78] & 2.46 (3.35) [-12.27, 20.83] \\

$\omega$B97X-D3BJ & 3.35 (4.57) [-11.02, 17.52]  & 2.56 (3.29) [-12.88, 14.51]  & 2.04 (2.74) [-7.39, 12.01]  & 2.48 (3.02) [-5.96, 7.80]   & 2.21 (2.98) [-8.04, 9.14]  & 2.36 (3.16) [-12.88, 17.52] \\

$\omega$B97M-V\tnote{d} & 2.70 (3.92) [-11.32, 14.93]  & 2.03 (2.74) [-13.43, 9.06]   & 1.74 (2.36) [-6.51, 12.83]  & 1.48 (1.99) [-5.16, 4.53]   & 1.62 (2.35) [-7.70, 6.11]  & 1.92 (2.65) [-13.43, 14.93] \\

$\omega$B97M-D3BJ\tnote{d} & 2.71 (3.73) [-11.17, 15.12]  & 1.86 (2.65) [-13.91, 7.07]   & 1.64 (2.21) [-7.25, 11.49]  & 1.42 (1.79) [-4.11, 4.02]   & 1.71 (2.28) [-7.08, 3.73]  & 1.81 (2.53) [-13.91, 15.12] \\

CAM-B3LYP & 3.25 (4.37) [-13.19, 16.21]  & 4.40 (6.18) [-32.27, 4.35]   & 2.44 (3.60) [-29.32, 11.32] & 2.01 (2.46) [-4.72, 5.40]   & 3.44 (4.12) [-10.97, 7.59] & 3.25 (4.77) [-32.27, 16.21] \\

\hline
& & & & & & \\

B2PLYP & 2.96 (4.58) [-15.82, 15.12]  & 3.44 (5.27) [-30.24, 5.63]   & 2.34 (3.70) [-29.70, 8.31]  & 1.92 (2.50) [-7.05, 3.88]   & 3.00 (3.92) [-12.40, 4.24] & 2.81 (4.41) [-30.24, 15.12] \\

B2PLYP-D3 & 2.49 (3.88) [-12.61, 15.06]  & 2.39 (3.58) [-20.07, 13.23]  & 1.88 (2.67) [-15.29, 9.37]  & 1.73 (2.35) [-5.91, 3.89]   & 2.45 (3.15) [-9.14, 3.61]  & 2.13 (3.16) [-20.07, 15.06] \\

mPW2PLYP-D & 2.34 (3.40) [-11.98, 13.89]  & 1.99 (2.96) [-15.63, 10.34]  & 1.68 (2.28) [-9.51, 9.22]   & 1.62 (2.07) [-4.70, 2.97]   & 2.29 (2.81) [-8.35, 3.78]  & 1.87 (2.67) [-15.63, 13.89] \\

\hline
& & & & & & \\

PM6 & 7.81 (11.04) [-40.57, 24.09] & 3.79 (5.36) [-11.99, 42.15]  & 3.10 (4.29) [-25.88, 19.44] & 4.35 (5.21) [-8.52, 9.92]   & 12.88 (16.28) [-41.87, 37.50] & 4.02 (6.10) [-41.87, 42.15] \\ 

PM7 & 9.34 (13.89) [-72.66, 32.44] &  3.44 (4.58) [-16.95, 23.05] & 2.83 (3.99) [-28.91, 15.79] & 6.04 (9.29) [-14.44, 35.40] & 11.93 (14.58) [-42.78, 25.76] & 3.89 (6.26) [-72.66, 35.40] \\
            \hline
\end{tabular}
}
\label{table:big}
\begin{tablenotes} 
\item[a] Nonhydrogens: 141, Hydrocarbons: 615, Subst. hydrocarbons: 849, Total: 1,674

\item[b] Nonhydrogens: 146, Hydrocarbons: 551, Subst. hydrocarbons: 805, Total: 1,571

\item[c] Nonhydrogens: 147, Hydrocarbons: 621, Total: 1,689

\item[d] Nonhydrogens: 139, Inorganic hydrides: 28, Total: 1,681
\end{tablenotes}
\end{threeparttable}
\end{table*}
\end{turnpage}

\section{Results and Discussions}
\subsection{Benchmark results for cWFTs and DFAs}

% \begin{figure}[htbp]
%     \centering
%     \includegraphics[width=8.6cm]{FIG_5.pdf}
%     \caption{Prediction errors in \hof~ across compound types and methods; MUE: mean unsigned error, SD: standard deviation, Max: maximum absolute deviation. For number of entries in each class of compounds and methods, see Table.~\ref{table:big}****}.
%   \label{fig:MaxSDMAD}
% \end{figure}

\begin{figure}[htbp]
    \centering
    \includegraphics[width=8.6cm]{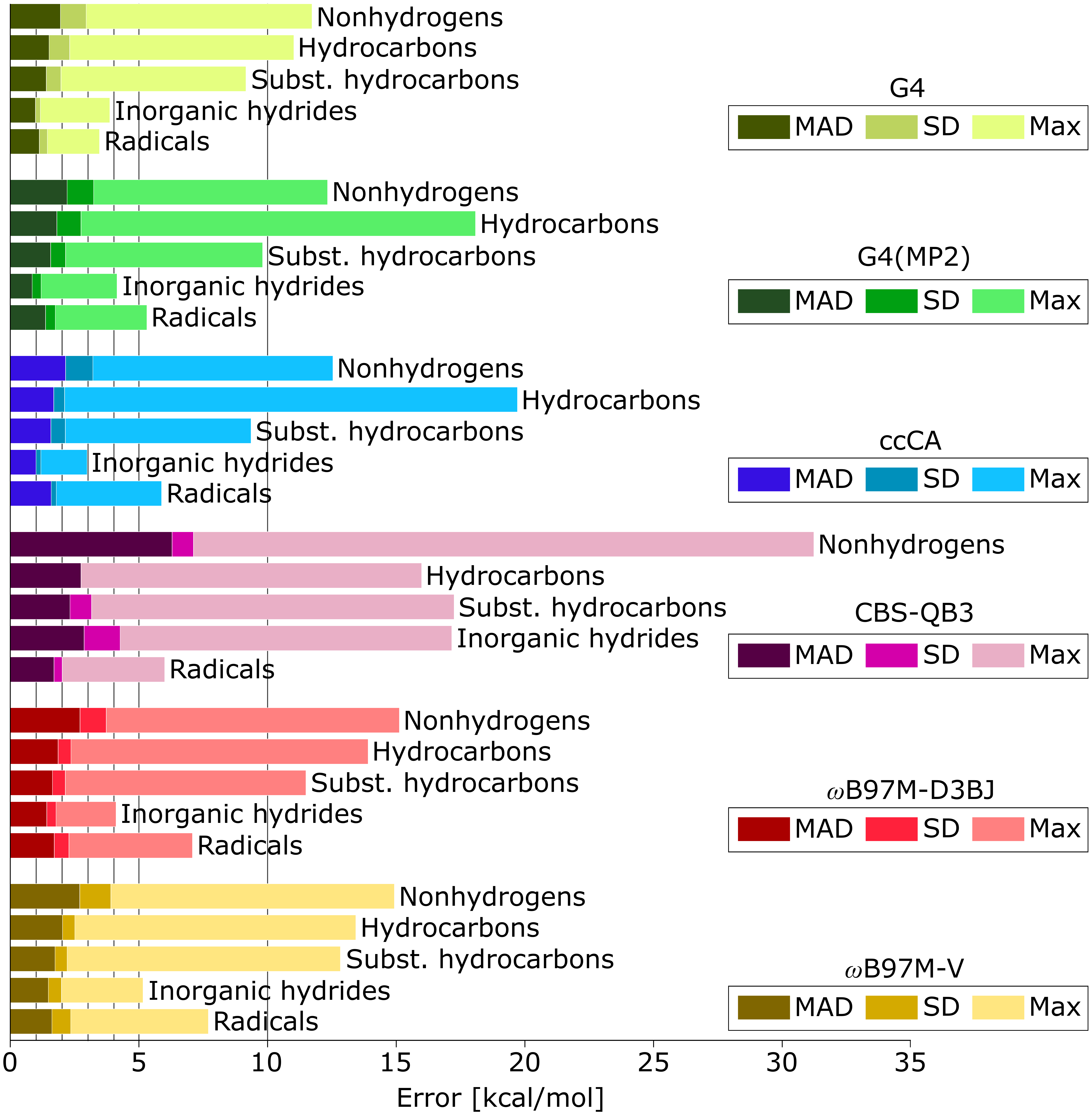}
    \caption{Prediction errors in \hof~ across compound types and methods; MUE: mean unsigned error, SD: standard deviation, Max: maximum absolute deviation. For number of entries in each class of compounds and methods, see Table.~\ref{table:big}}.
   \label{fig:MaxSDMAD}
\end{figure}

\begin{figure}[ht]
    \centering
    \includegraphics[width=8.6cm]{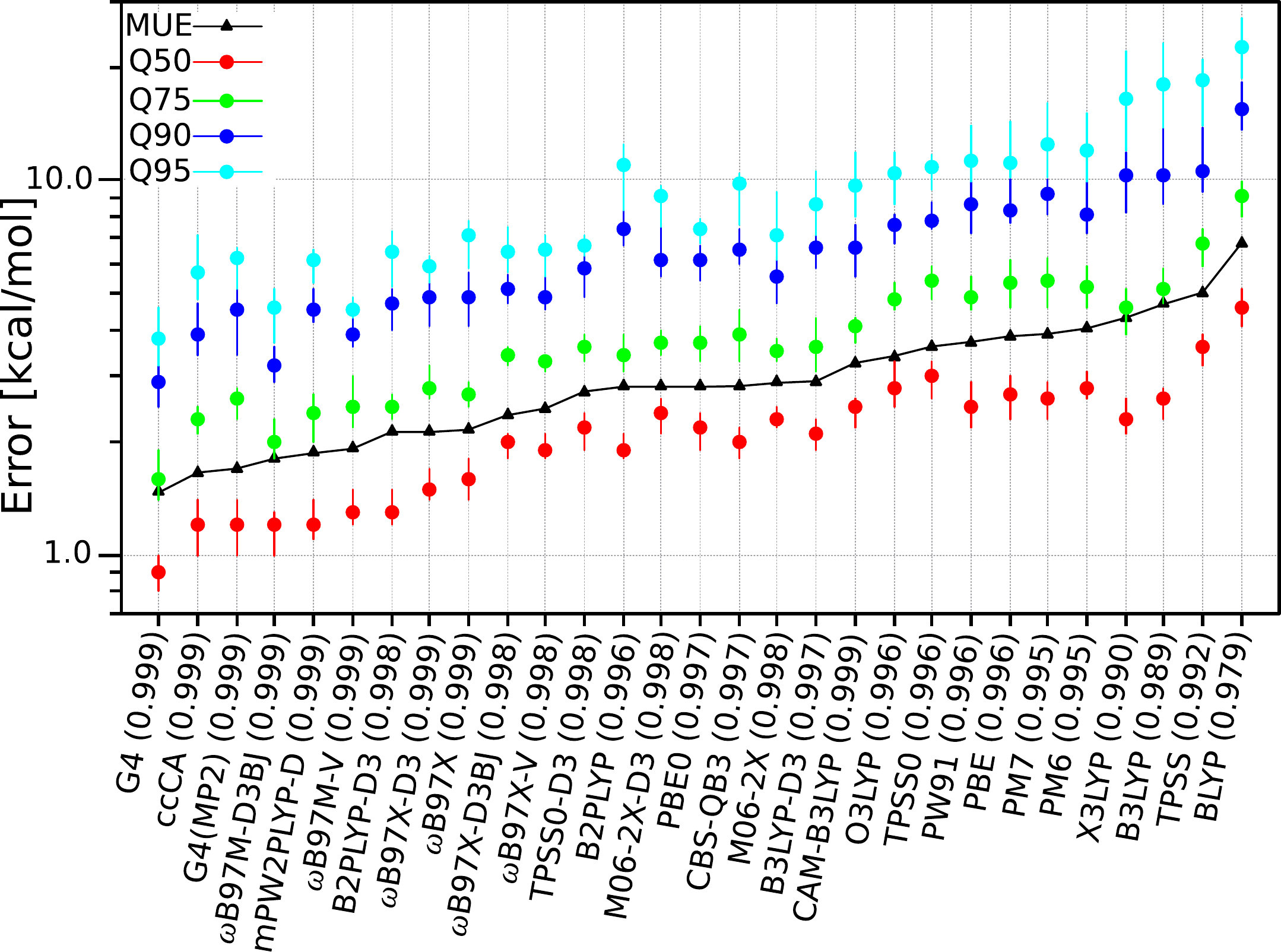}
    \caption{Ranking of the prediction accuracy of cWFTs, DFAs and semi-empirical models. For each method, the mean unsigned error (MUE) and cumulative probability percentiles, $Qn\,(n=50,75,90,\,{\mathrm {and}}\,95)$ are reported in log-scale. Methods are sorted in ascending order of MUE. Values in square brackets are the Spearman rank correlations w.r.t experimental values.
    % \INGREEN{There are slight errors in the names of G4MP2 and mPW2PLYP}
    Vertical lines on the percentile points denote the standard deviation estimated via
    bootstrapping.
    }
    \label{fig:NeNa}
\end{figure}

\begin{figure}[htbp]
    \centering
    \includegraphics[width=8.6cm]{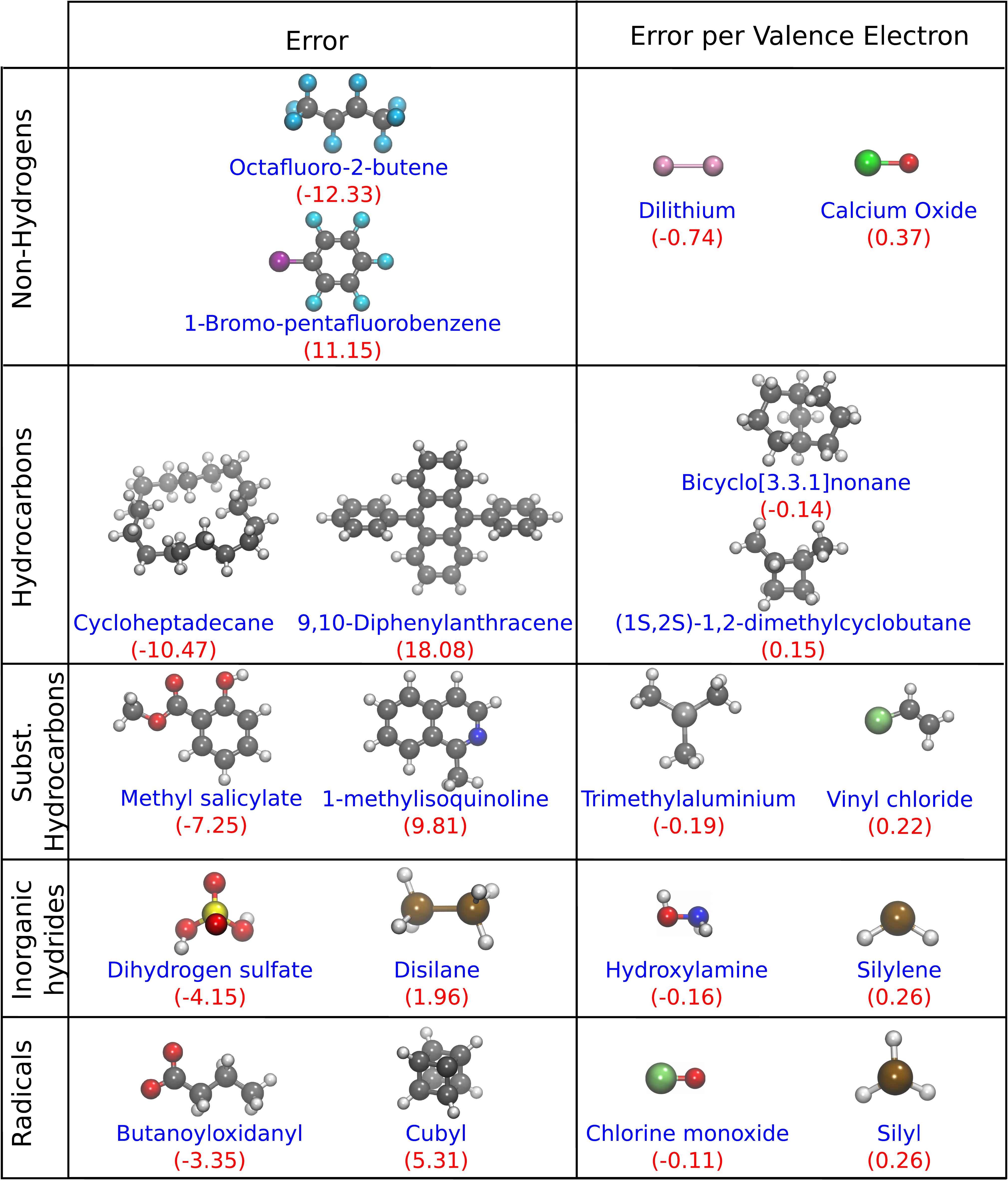}
    \caption{Left column presents molecules in the PPE1694 dataset exhibiting the
    highest deviations in G4(MP2) predicted \hof. Right column shows molecules with extreme errors per valence electron.}
    \label{fig:extremes}
\end{figure}

\begin{table}[ht]
\caption{Comparison between errors in G4(MP2), ccCA and G4 for G3/05 and Pedley CHONF. Both results from previous and current studies are included for all classes of compounds.}
\begin{threeparttable}
{\begin{tabular}{ l l l l}
\hline
Dataset                    & G4(MP2)                  & ccCA                     & G4 \\
\hline
A. G3/05 (452)\tnote{a}                        & 1.05, 1.04\tnote{b}~~~~~~& 0.98, 0.99\tnote{c}~~~~~~& 0.83\tnote{d}\\
1. $\Delta H_f^\circ$ (270)           & 1.00, 0.99\tnote{b}& 0.93, 0.95\tnote{c}& 0.80\tnote{d}\\
 ~~~~ -Nonhydrogens (79)             & 1.46, 1.44\tnote{b}& 1.04 & 1.13\tnote{d}\\
 ~~~~ -Hydrocarbons (38)              & 0.64, 0.63\tnote{b}& 0.96 & 0.48\tnote{d}\\
 ~~~~ -Subst. Hydrocarbons (100)         & 0.84, 0.83\tnote{b}& 0.85 & 0.68\tnote{d}\\
 ~~~~ -Inorganic hydrides (19)        & 0.92, 0.94\tnote{b}& 0.99 & 0.92\tnote{d}\\
 ~~~~ -Radicals (34)                  & 0.83, 0.86\tnote{b}& 0.82 & 0.66\tnote{d}\\
 2. IP (103)\tnote{a}                          & 1.08, 1.07\tnote{b}& 1.08, 1.09\tnote{c}& 0.91\tnote{d}\\
 ~~~~ -Atoms (26)                     & 1.12, 1.13\tnote{b}& 0.55               & 0.65\tnote{d}\\
 ~~~~ -Molecules (77)                 & 1.06, 1.05\tnote{b}& 1.26               & 0.99\tnote{d}\\
 3. EA (63)                           & 1.26, 1.23\tnote{b}& 0.97, 1.03\tnote{c}& 0.83\tnote{d}\\
 ~~~~ -Atoms (14)                     & 1.86, 1.84\tnote{b}& 0.89               & 0.91\tnote{d}\\
 ~~~~ -Molecules (49)                 & 1.10, 1.06\tnote{b}& 1.00               & 0.81\tnote{d}\\
 4. PA (10)                           & 0.66, 0.67\tnote{b}& 1.23, 0.93\tnote{c}& 0.84\tnote{d} \\
 5. BE (6)                            & 1.29, 1.28\tnote{b}& 1.15, 0.58\tnote{c}& 1.12\tnote{d} \\
 %II. Pedley, $\Delta H_f^\circ$ (459) & 0.78, 0.79\tnote{e}& 1.01& \\
 %~~~~ -Hydrocarbons (175)             & 0.61, 0.68\tnote{e}& 0.88& \\
 %~~~~ -Subst. hydrocarbons (284)         & 0.88, 0.86\tnote{e}& 1.08& \\
 B. Pedley CHONF, $\Delta H_f^\circ$ (459) & 0.76, 0.79\tnote{e}& 1.01 & 0.69 \\
 ~~~~ -Hydrocarbons (175)             & 0.61, 0.68\tnote{e}& 0.84 & 0.51 \\
 ~~~~ -Subst. hydrocarbons (284)         & 0.86, 0.86\tnote{e}& 1.12 & 0.80\\ 
 
\hline
\end{tabular}}
\label{table:small}
\begin{tablenotes} 
\item[a] We have excluded two triplet entries for H$_2$S and N$_2$
\item[b] from \RRef{curtiss2007gaussian_a}
\item[c] from \RRef{deyonker2009towards} %see Table 8
\item[d] from \RRef{curtiss2007gaussian}
\item[e] from \RRef{narayanan2019accurate}
\end{tablenotes}
\end{threeparttable}
\end{table}

While cWFTs are the recommended choice for modeling \hof, the favorable accuracy-to-speed trade-offs of DFAs have facilitated $\Delta H_f^\circ$ predictions for large molecules through isodesmic reaction schemes\cite{jaidann2010dft,turker2010dft} and group additivity schemes\cite{guthrie2001heats}. A benchmark across popular DFAs on a  curated dataset such as PPE1694 (see Fig.~\ref{fig:datasets}) may provide insights
into how the performances of these methods
can be refined through proper selection of basis sets and additional dispersion corrections. To this end, we embark upon comprehensive benchmarking of G4, G4(MP2), ccCA, CBS-QB3, 23 popular DFAs, as well as the semi-empirical methods, PM6 and PM7. 

Table~\ref{table:big} summarizes MUEs, RMSEs and maximum signed errors for each method across various classes of compounds. It is apparent while moving across 
the subsets that the performance of cWFTs---G4, G4(MP2), and ccCA---are consistent irrespective of the compound type, with total MUEs $\le 1.70$ kcal/mol.
CBS-QB3 performs poorly relative to other cWFTs, notably for non-hydrogens with an MUE over 6~kcal/mol.

We note the GGA functionals to consistently under-perform across all subsets. A closer look reveals errors to be mostly systematic as larger MUEs are usually encountered for subsets with greater structural complexities. Hybrid GGAs \INBLUE{mostly} improve upon GGA \INBLUE{with PBE0 showing the best performance.} 
\INBLUE{This is because hybrid GGAs are designed to correct for the spurious self-interaction error in semi-local DFAs\cite{rosch1997comment,tozer2005computation,ramakrishnan2009dft+}}. Overall, as we move up the \textit{Jacob's ladder}\cite{perdew2001jacob}, we observe a consistent drop in MUE. \INBLUE{Most notably, we find the performance of long-range tuned hybrid DFAs on par with double hybrid ones. 
The importance of capturing long-range effects can be further understood by noting that explicit inclusion of an empirical dispersion correction overall improves the performance of most DFAs. Semi-empirical methods PM6 and PM7 under-perform across all subsets except for hydrocarbons
and substituted hydrocarbons, where their accuracy is on par with long-range tuned functionals with PM7 slightly outperforming PM6.}

% \INGREEN{Most notably, we find the long-range tuned hybrid DFAs to perform evenly with the double hybrid functionals. The importance of capturing long-range effects can be further understood by noting that explicit inclusion of an empirical dispersion correction mostly improves the performance across DFAs. Semi-empirical methods PM6 and PM7 under-perform across all subsets except for hydrocarbons and substituted hydrocarbons, where their accuracy is on par with many functionals, with PM7 slightly outperforming PM6.}

For the Top-6 methods: G4, G4(MP2), ccCA, CBS-QB3, \INBLUE{$\omega$B97M-D3BJ, and $\omega$B97M-V}, we have graphically summarized the error metrics in Fig.~\ref{fig:MaxSDMAD}. As noted in Table.~\ref{table:big}, G4 ranks best followed by G4(MP2) and ccCA, both exhibiting comparable accuracies. In terms of the maximum absolute deviation (Max), ccCA seems slightly better than G4 and G4(MP2) for inorganic hydrides. \INBLUE{Barring radicals, $\omega$B97M-D3BJ performs slightly better than $\omega$B97M-V across the subsets. Both of them significantly outperform CBS-QB3.}

A drawback of relying on mean-based error metrics is that, these do not provide a complete picture of the error distribution. For this purpose, percentile-based metrics have been shown to be more suitable (see \RRef{pernot2018probabilistic} and references therein). Fig.~\ref{fig:NeNa} presents mean- and percentile-based metrics for all the computational methods studied here. A method with good prediction accuracy should show smaller MUEs as well as small values of $Q_N$, where $N > 50$. Spearman rank correlation ($\rho$) between two sets is a good indicator for qualitative agreement between them\cite{spearmancoeff}. 
Overall, methods with low
$Q_N$ or MUE, show $\rho \approx 1$. In general, we find dispersion corrections to improve the $\rho$ of DFAs.

%Compared to 1,694 experimental \hof~ values, we find G4, G4(MP2), ccCA to show a strong correlation amounting to $\rho=0.999$.  \INBLUE{This is closely followed by $\omega$B97M-D3BJ, mPW2PLYP-D and $\omega$B97M-V, all with similar $\rho$.}

%As noted for MUE, the percentile metrics and $\rho$ of PM6 and PM7 are consistently superior than that of \INBLUE{many} GGA methods. \INGREEN{"Better to avoid the last statement"} This is because, these simple methods have been trained to deliver accurate values of \hof~ for hydrocarbons and substituted hydrocarbons, both sets constituting a large fraction of PPE1694.

\begin{figure*}[ht]
    \centering
    \includegraphics[width=13.5cm]{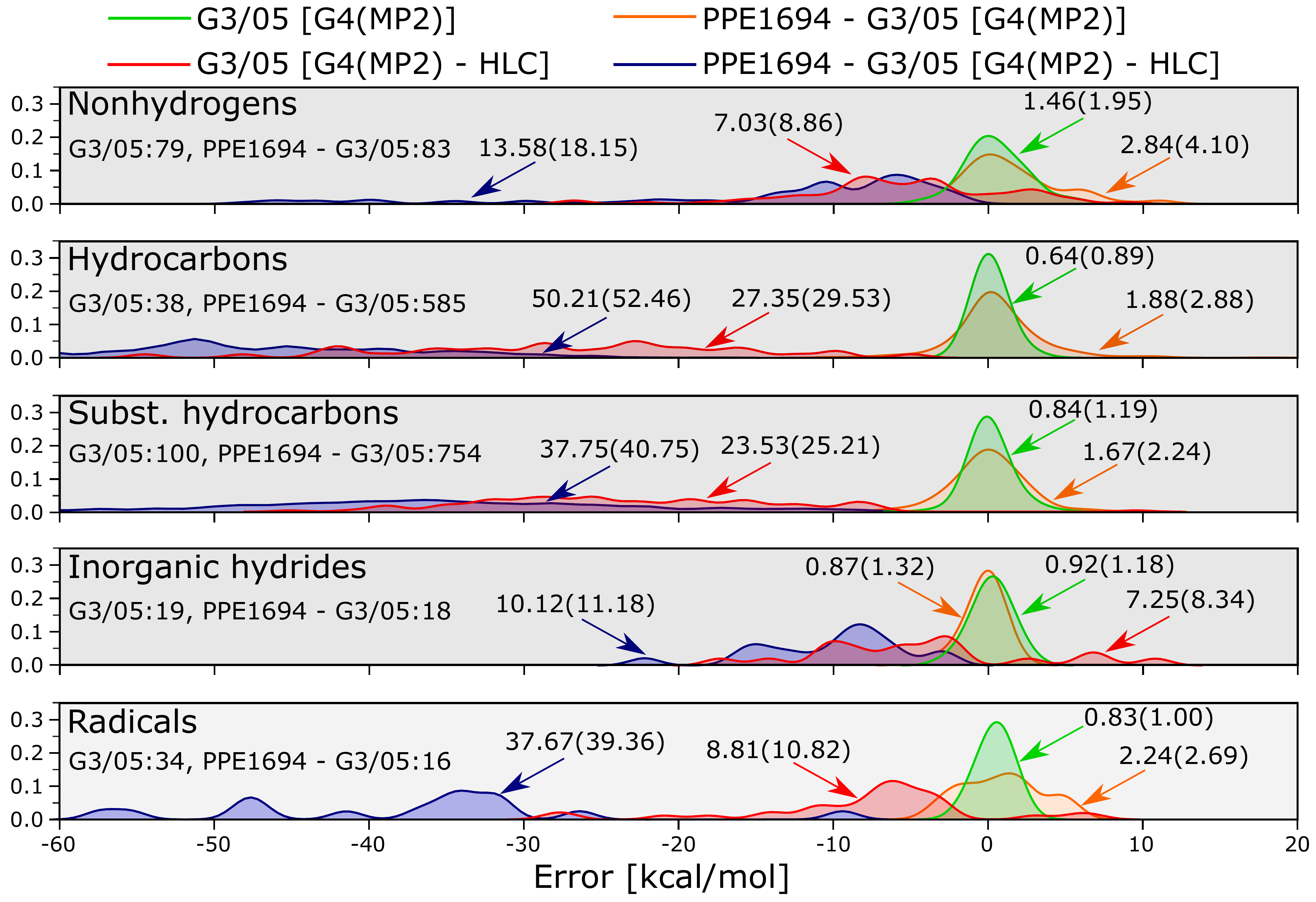}
    \caption{
    Influence of HLC in determining \hof~for various classes of compounds in the G3/05 dataset
    and rest of the compounds in the PPE1694 dataset. For both sets, results are shown separately with and without the HLC term. The arrows point to MUE (RMSE) for each error distribution. The lower bound of the error is set to -60 kcal/mol for clarity.}
    \label{fig:alexbreakdown1}
\end{figure*}

An examination of the compounds showing extreme ({\it i.e.} most-positive or most-negative) errors for a given computational method often reveals if the error is systematic or non-systematic in nature. 
For the latter to be evident one has to consider normalized or intensive errors such as error-per-electron or error-per-atom. In the case of G4(MP2), along with the total error, we also consider error-per-valence-electron and inspected those compounds exhibiting extreme errors (see Fig.~\ref{fig:extremes}). Barring hydrocarbons, we find the compounds with extreme intensive and extensive ({\it i.e.} unnormalized) error to mostly comprise heavy atoms. Predominant of the compounds with large intensive error are with fewer heavy atoms, while those with large extensive (or unnormalized) error consist of either several atoms or several electrons. The hydrocarbons cycloheptadecane and 9,10-diphenylanthracene show the largest extensive error; from opposite signs of the errors of these two compounds, one may speculate that the source of their errors cannot be due to the uncertainties associated with the parameters used in the enthalpy evaluation (see Section \ref{sec:sys} for a discussion). 

It is important to compare the accuracy of the G4(MP2) and ccCA results presented above, based on the present implementation, to that of legacy implementations of these cWFTs. Firstly, all our calculations are based on a framework employing spherical primitive Gaussian type
orbitals (GTOs), while Pople basis sets---used for B3LYP geometry relaxation, and G4(MP2) energies---are conventionally used in the Cartesian primitive GTO framework. Secondly, the ccCA-P\cite{deyonker2009towards} formalism uses B3LYP/cc-pVTZ reference geometry with Hartree-Fock and MP2 energies extrapolated separately to the CBS limit, relativistic DKH2 corrections for open-shell molecules calculated with a spin-collinear ({\it i.e.} UHF) reference wavefunction, and employs a different scale factor for ZPVE. To this end, in Table~\ref{table:small}, we compare prediction errors in our cWFT calculations for the entire G3/05 dataset to that of previously published results with same methods. For comparison, we have also summarized previous results based on the more accurate G4 theory. 

As far as the G4(MP2) results are concerned, going from Cartesian GTOs to spherical GTOs leads to a tiny increase in the MUE by 0.01 kcal/mol. 
Our prediction error for ccCA fares rather well when compared to ccCA-P values from \RRef{deyonker2009towards}, while the original formalism---with a larger basis set for geometry optimization---showing much smaller error for the binding energy of hydrogen bonded dimers. 
While we believe that the accuracy of our ccCA results can improve 
when adopting the ccCA settings to be similar to that of the 
ccCA-P\cite{deyonker2009towards} implementation, here 
we used settings that render a comparison between G4(MP2) and ccCA seamless.

Further, we have extended the comparison between cWFTs to the slightly larger dataset -- Pedley CHONF (see Section\ref{sec:dataset}); we note our G4(MP2) results to agree fairly well with
that of previously reported values from \RRef{narayanan2019accurate}. 
Interestingly, for G4 and G4(MP2) we find the accuracies for hydrocarbons 
and their substituted analogues 
to be retained when going from the G3/05 set to Pedley CHONF.  
As for ccCA, the MUE of 1.01 kcal/mol for the Pedley CHONF dataset is 
comparable to that of the total
error for the G3/05 set, indicating the small deviations in both values to lie within the uncertainties in the reference experimental values. 

\subsection{Transferability of HLC in G4(MP2)}

An interesting point of concern in the G$n$-series of cWFTs---more specifically G4(MP2)---is the role HLC plays in the model. 
The total G4(MP2) electronic energy takes the form
\begin{eqnarray}
E_0^\mathrm{G4(MP2)} & = & E_{\mathrm{GTBAS1}}^{\mathrm{CCSD(T)}}  + \Delta E^{\mathrm{MP2}} + \Delta E^{\mathrm{HF}} + \nonumber \\
& & \mathrm{ZPVE} + \mathrm{SO} + \mathrm{HLC}
\label{eq:g4mp2}
\end{eqnarray}

The HLC term, as pointed out by Martin\cite{martin1997coupling,martin1996ab,martin1992performance,martin1994performance,martin2005computational}, accounts for the residual error introduced in the additive model when there is significant coupling between the one-particle basis set and $N$-electron correlated wavefunction. However, it is long known that the success of the HLC correction in G$n$ methods is strongly coupled to the choice of basis sets employed. To quote John Pople\cite{pople1989gaussian} ---from the first study on the G1 method---{\it ``Uniform application of the HLC is only sensible if the basis used, 6--311+G$^{**}$(2df), is reasonably balanced, meaning that residual errors per electron are approximately constant over a wide range of molecules. ''}. HLC has been modified over the years through G1\cite{pople1989gaussian}, G2\cite{curtiss1991gaussian}, G3\cite{curtiss1998gaussian_a} and
G4\cite{curtiss2007gaussian} studies.
In G4(MP2)\cite{curtiss2007gaussian_a}, the HLC terms 
take the same form as in G4
but with parameters optimized separately:
\begin{equation}
    \mathrm{HLC^{G4(MP2)}} =
    \begin{cases}
      -An_\beta  \\
      -A^{'}n_\beta - B(n_\alpha - n_\beta) \\
      -Cn_\beta - D(n_\alpha - n_\beta) \\
       -E 
   \end{cases}
   \label{ref:hlc}
\end{equation}
where the terms bear the same meaning as in the original G4 study\cite{curtiss2007gaussian}.

\begin{table*}[ht]
    \centering
    \caption{Accuracies of G4, G4(MP2), ccCA, CBS-QB3, $\omega$B97M-D3BJ and $\omega$B97M-V for the prediction of isomerization enthalpies for 32 sets of constitutional isomers: MUE is mean unsigned error (in kcal/mol), RMSE is root-mean-square-error (in kcal/mol), and $\rho$ is the Spearman rank correlation coefficient; the latter quantity is reported per stoichiometry.
    }
    {
    \begin{tabular}{l l l l l l l}
    \hline
    Stoichiometry (\#)~~~~~~& \multicolumn{6}{l}{MUE (RMSE, $\rho$)}\\
    \cline{2-7} 
            &  \multicolumn{1}{l}{G4}~~~~& \multicolumn{1}{l}{G4(MP2)}& \multicolumn{1}{l}{ccCA}& \multicolumn{1}{l}{CBS-QB3} & \multicolumn{1}{l}{$\omega$B97M-D3BJ} & \multicolumn{1}{l}{$\omega$B97M-V}\\
            
\hline
C$_{4}$H$_{6}$(7)         &0.91 (1.13, 0.98)~~~~ & 1.03 (1.18, 0.98)~~~~ & 0.80 (1.02, 1.00)~~~~ & 0.96 (1.21, 0.98)~~~~ & 1.50 (2.12, 0.98)~~~~ & 2.19 (2.68, 1.00)  \\
C$_{5}$H$_{8}$(12)        &0.81 (0.89, 0.99) & 0.85 (0.94, 0.98) & 0.40 (0.60, 0.99) & 0.51 (0.61, 0.99) & 1.02 (1.22, 0.98) & 1.34 (1.56, 0.98)  \\
C$_{5}$H$_{10}$(11)       &0.79 (0.86, 0.99) & 0.86 (0.90, 0.99) & 0.26 (0.37, 0.99) & 0.52 (0.64, 0.99) & 0.42 (0.58, 0.99) & 0.88 (0.90, 0.99)  \\
C$_{6}$H$_{8}$(8)         &1.11 (1.18, 0.98) & 1.14 (1.24, 0.98) & 0.89 (1.03, 1.00) & 1.05 (1.13, 0.98) & 0.72 (0.81, 1.00) & 1.10 (1.23, 1.00)  \\
C$_{6}$H$_{10}$(26)       &0.96 (1.47, 0.99) & 1.02 (1.54, 0.99) & 0.91 (1.35, 0.99) & 0.92 (1.38, 0.99) & 1.15 (1.45, 0.99) & 1.42 (1.70, 0.98)  \\
C$_{6}$H$_{12}$(26)       &0.87 (1.59, 0.97) & 0.88 (1.55, 0.96) & 0.84 (1.43, 0.97) & 0.84 (1.45, 0.97) & 1.02 (1.61, 0.97) & 1.57 (1.93, 0.97)  \\
C$_{7}$H$_{8}$(9)         &1.79 (2.01, 0.96) & 1.64 (1.96, 0.96) & 1.63 (1.88, 0.96) & 0.94 (1.20, 0.96) & 1.30 (1.46, 0.96) & 2.21 (2.63, 0.96)  \\
C$_{7}$H$_{12}$(31)       &1.18 (1.50, 0.97) & 1.30 (1.64, 0.97) & 1.38 (1.68, 0.97) & 1.48 (1.74, 0.97) & 1.36 (1.68, 0.98) & 1.95 (2.37, 0.98)  \\
C$_{7}$H$_{14}$(39)       &0.66 (0.86, 0.91) & 0.70 (0.94, 0.91) & 0.60 (0.77, 0.93) & 0.65 (0.83, 0.92) & 0.49 (0.66, 0.94) & 0.93 (1.12, 0.94)  \\
C$_{7}$H$_{16}$(8)        &0.37 (0.62, 0.97) & 0.37 (0.62, 0.98) & 0.47 (0.62, 0.97) & 0.38 (0.61, 0.97) & 0.62 (0.77, 0.95) & 0.57 (0.69, 0.95)  \\
C$_{8}$H$_{10}$(15)       &1.43 (1.73, 0.98) & 1.41 (1.78, 0.98) & 1.34 (1.64, 0.98) & 1.33 (1.60, 0.98) & 1.13 (1.46, 0.99) & 1.71 (2.11, 0.99)  \\
C$_{8}$H$_{14}$(14)       &1.14 (2.25, 0.94) & 1.18 (2.28, 0.95) & 1.07 (2.19, 0.93) & 1.10 (2.13, 0.94) & 1.17 (2.24, 0.94) & 1.29 (2.10, 0.94)  \\
C$_{8}$H$_{16}$(77)       &4.81 (4.98, 0.89) & 5.04 (5.21, 0.88) & 4.60 (4.76, 0.89) & 4.54 (4.73, 0.89) & 4.87 (5.01, 0.89) & 3.99 (4.19, 0.89)  \\
C$_{8}$H$_{18}$(17)       &1.77 (2.17, 0.69) & 1.83 (2.22, 0.69) & 1.23 (1.60, 0.59) & 1.56 (1.98, 0.72) & 1.24 (1.38, 0.53) & 1.21 (1.37, 0.56)  \\
C$_{9}$H$_{10}$(7)        &0.66 (0.80, 0.86) & 0.75 (0.89, 0.86) & 0.82 (0.94, 0.76) & 0.90 (1.13, 0.93) & 0.82 (0.86, 0.76) & 1.50 (1.66, 0.76)  \\
C$_{9}$H$_{12}$(13)       &0.82 (1.04, 0.99) & 1.20 (1.68, 0.99) & 1.11 (1.41, 0.99) & 1.15 (1.50, 1.00) & 0.89 (0.99, 0.99) & 0.72 (1.07, 0.99)  \\
C$_{9}$H$_{16}$(9)        &2.91 (3.63, 0.93) & 2.96 (3.71, 0.93) & 2.42 (3.57, 0.96) & 2.89 (3.56, 0.94) & 2.39 (3.43, 0.95) & 2.42 (3.23, 0.95)  \\
C$_{9}$H$_{18}$(11)       &3.62 (3.91, 0.71) & 3.59 (3.90, 0.71) & 3.43 (3.78, 0.62) & 3.55 (3.85, 0.71) & 2.95 (3.38, 0.72) & 3.07 (3.45, 0.73)  \\
C$_{10}$H$_{10}$(10)      &5.35 (5.91, 1.00) & 5.51 (6.21, 1.00) & 5.66 (6.17, 1.00) & 5.97 (6.63, 1.00) & 6.26 (6.93, 1.00) & 6.44 (7.22, 1.00)  \\
C$_{10}$H$_{16}$(9)       &2.61 (3.14, 0.99) & 2.32 (2.89, 0.99) & 2.81 (3.15, 0.99) & 3.01 (3.49, 0.99) & 2.27 (2.60, 0.99) & 3.34 (3.67, 0.99)  \\
C$_{7}$H$_{9}$N(13)       &1.08 (1.70, 0.99) & 1.10 (1.98, 0.99) & 1.23 (2.02, 0.98) & 1.32 (2.14, 0.99) & 1.38 (2.52, 0.99) & 1.30 (2.43, 0.99)  \\
C$_{5}$H$_{10}$O(11)      &0.50 (0.64, 0.99) & 0.45 (0.56, 0.99) & 0.56 (0.71, 0.98) & 0.71 (0.97, 0.97) & 0.86 (0.94, 0.97) & 0.75 (0.94, 0.97)  \\
C$_{5}$H$_{12}$O(11)      &0.64 (1.18, 0.93) & 0.66 (1.24, 0.91) & 0.68 (1.20, 0.93) & 0.74 (1.21, 0.93) & 0.85 (1.23, 0.95) & 0.75 (1.16, 0.92)  \\
C$_{6}$H$_{12}$O(7)       &0.71 (0.75, 0.90) & 0.60 (0.68, 0.93) & 0.76 (0.90, 0.86) & 0.84 (0.92, 0.90) & 0.57 (0.74, 0.90) & 0.90 (1.13, 0.86)  \\
C$_{6}$H$_{14}$O(7)       &0.73 (0.80, 0.83) & 0.69 (0.76, 0.95) & 0.84 (0.91, 0.98) & 0.71 (0.86, 0.81) & 0.60 (0.74, 0.95) & 0.57 (0.76, 0.95)  \\
C$_{8}$H$_{10}$O(12)      &1.37 (1.60, 0.94) & 1.37 (1.59, 0.94) & 1.49 (1.69, 0.92) & 1.38 (1.62, 0.92) & 1.61 (1.85, 0.90) & 1.56 (1.82, 0.92)  \\
C$_{4}$H$_{8}$O$_{2}$(9)  &1.49 (1.76, 1.00) & 1.43 (1.64, 1.00) & 1.40 (1.57, 1.00) & 1.92 (2.25, 1.00) & 1.15 (1.53, 1.00) & 1.65 (1.89, 1.00)  \\
C$_{4}$H$_{10}$O$_{2}$(8) &2.68 (2.88, 1.00) & 2.60 (2.76, 1.00) & 2.30 (2.47, 1.00) & 2.84 (3.16, 1.00) & 2.82 (2.96, 1.00) & 2.88 (3.06, 1.00)  \\
C$_{5}$H$_{10}$O$_{2}$(15)&3.09 (3.45, 0.98) & 2.78 (3.14, 0.98) & 2.68 (3.05, 0.98) & 3.71 (4.13, 0.98) & 2.31 (2.57, 0.98) & 2.99 (3.41, 0.98)  \\
C$_{6}$H$_{12}$O$_{2}$(16)&2.07 (2.79, 0.95) & 2.28 (2.96, 0.95) & 2.70 (3.36, 0.95) & 1.82 (2.65, 0.95) & 3.31 (3.89, 0.94) & 2.53 (3.24, 0.94)  \\
C$_{5}$H$_{12}$S(9)       &0.61 (0.70, 0.87) & 0.64 (0.78, 0.88) & 0.45 (0.64, 0.88) & 0.67 (0.78, 0.89) & 0.64 (0.79, 0.83) & 0.53 (0.68, 0.83)  \\
C$_{6}$H$_{14}$S(8)       &1.29 (1.83, 0.87) & 1.31 (1.85, 0.87) & 0.87 (1.57, 0.78) & 1.25 (1.79, 0.85) & 0.96 (1.53, 0.82) & 0.85 (1.48, 0.80)  \\
\hline
Total(485)                &1.94 (2.75) & 2.00 (2.85) & 1.86 (2.67) & 1.94 (2.75) & 1.96 (2.79) & 2.04 (2.72)  \\
 \hline
\end{tabular}}
    \label{tab:isomers}
\end{table*}

\begin{table*}[ht]
    \centering
    \caption{Accuracies of methods in determining the \hof~ of global minimum of each stoichiometry in ISO32****}
    \resizebox{\linewidth}{!}{
    \begin{tabular}{r l r r r r r r r}
    %\begin{tabular}{r l{4cm} r{4cm} r{4cm} r{4cm} r{4cm} r{4cm} r{4cm} r{4cm}}
    \hline
    \#~~~~&
    Global minimum (Stoichiometry)~~~~~~&$\Delta H_f^{\circ,{\rm exp.}}$ &\multicolumn{6}{l}{$\Delta H_f^{\circ,{\rm calc.}}$  ($\Delta H_f^{\circ,{\rm exp.}}-\Delta H_f^{\circ,{\rm calc.}}$) }\\
    \cline{4-9} 
        &    &  & \multicolumn{1}{l}{G4}& \multicolumn{1}{l}{G4(MP2)}& \multicolumn{1}{l}{ccCA}& \multicolumn{1}{l}{CBS-QB3} & \multicolumn{1}{l}{$\omega$B97M-D3BJ} & \multicolumn{1}{l}{$\omega$B97M-V}\\
            
\hline
1&1,3-Butadiene (C$_4$H$_6$)                                    &    26.23  &    26.52 ( -0.29) &    25.72 (  0.51) &   27.21 (  -0.98) &   28.27 (  -2.04) &   27.42 (  -1.19) &   28.24 (  -2.01) \\  
2&Cyclopentene (C$_5$H$_8$)                                     &     7.91  &     8.94 ( -1.03) &     8.44 ( -0.53) &    8.92 (  -1.01) &   10.35 (  -2.44) &    8.63 (  -0.72) &    8.50 (  -0.59) \\
3&Cyclopentane (C$_5$H$_{10}$)                                  &   -18.28  &   -17.39 ( -0.89) &   -17.36 ( -0.92) &  -17.87 (  -0.41) &  -17.00 (  -1.28) &  -17.73 (  -0.55) &  -18.00 (  -0.28) \\
4&Methylcyclopentadiene (C$_6$H$_8$)                            &    23.90  &    23.68 (  0.22) &    22.88 (  1.02) &   24.63 (  -0.73) &   25.96 (  -2.06) &   23.94 (  -0.04) &   23.80 (   0.10) \\
5&Cyclohexene (C$_6$H$_{10}$)                                   &    -1.15  &    -0.66 ( -0.49) &    -1.14 ( -0.01) &   -0.38 (  -0.77) &    1.14 (  -2.29) &   -0.78 (  -0.37) &   -0.84 (  -0.31) \\
6&Cyclohexane (C$_6$H$_{12}$)                                   &   -29.49  &   -28.89 ( -0.60) &   -28.86 ( -0.63) &  -28.99 (  -0.50) &  -27.50 (  -1.99) &  -28.73 (  -0.76) &  -28.97 (  -0.52) \\
7&Toluene (C$_7$H$_8$)                                          &    12.00  &    11.98 (  0.02) &    11.02 (  0.98) &   13.59 (  -1.59) &   13.60 (  -1.60) &   11.35 (   0.65) &   11.62 (   0.38) \\
8&Bicyclo[2.2.1]heptane (C$_7$H$_{12}$)                         &   -12.96  &   -13.47 (  0.51) &   -13.58 (  0.62) &  -12.87 (  -0.09) &  -11.91 (  -1.05) &  -12.30 (  -0.66) &  -13.69 (   0.73) \\
9&Methylcyclohexane (C$_7$H$_{14}$)                             &   -36.98  &   -36.64 ( -0.34) &   -36.60 ( -0.38) &  -36.44 (  -0.54) &  -34.85 (  -2.13) &  -35.85 (  -1.13) &  -36.21 (  -0.77) \\
10&2,2-Dimethylpentane (C$_7$H$_{16}$)                          &   -49.20  &   -49.27 (  0.07) &   -49.11 ( -0.09) &  -49.00 (  -0.20) &  -47.49 (  -1.71) &  -47.46 (  -1.74) &  -47.23 (  -1.97) \\
11&1,3-Dimethylbenzene (C$_8$H$_{10}$)                          &     4.12  &     4.05 (  0.07) &     3.24 (  0.88) &    6.13 (  -2.01) &    6.05 (  -1.93) &    3.92 (   0.20) &    4.12 (  -0.00) \\
12&Bicyclo[2.2.2]octane (C$_8$H$_{14}$)                         &   -23.66  &   -22.78 ( -0.88) &   -22.81 ( -0.85) &  -21.98 (  -1.68) &  -20.74 (  -2.92) &  -21.61 (  -2.05) &  -22.95 (  -0.71) \\
13&(1R,2R,3S)-1,2,3-trimethylcyclopentane (C$_8$H$_{16}$)       &   -45.09  &   -40.20 ( -4.89) &   -40.09 ( -5.00) &  -39.59 (  -5.50) &  -37.99 (  -7.10) &  -38.66 (  -6.43) &  -39.24 (  -5.85) \\
14&2,2,3,3-Tetramethylbutane (C$_8$H$_{18}$)                    &   -53.92  &   -55.01 (  1.09) &   -54.92 (  1.00) &  -53.82 (  -0.10) &  -52.74 (  -1.18) &  -51.57 (  -2.35) &  -51.53 (  -2.39) \\
15&2,3-Dihydro-1H-indene (C$_9$H$_{10}$)                        &    14.51  &    13.90 (  0.61) &    12.96 (  1.55) &   16.21 (  -1.70) &   15.82 (  -1.31) &   13.63 (   0.88) &   13.05 (   1.46) \\
16&1,3,5-Trimethylbenzene (C$_9$H$_{12}$)                       &    -3.80  &    -3.92 (  0.12) &    -4.58 (  0.78) &   -1.41 (  -2.39) &   -1.41 (  -2.39) &   -3.56 (  -0.24) &   -3.44 (  -0.36) \\
17&trans-Octahydro-1H-indene (C$_9$H$_{16}$)                    &   -31.43  &   -31.48 (  0.05) &   -31.51 (  0.08) &  -30.67 (  -0.76) &  -29.01 (  -2.42) &  -30.26 (  -1.17) &  -31.49 (   0.06) \\
18&(1a,3a,5a)-1,3,5-Trimethylcyclohexane (C$_9$H$_{18}$)        &   -50.69  &   -52.19 (  1.50) &   -52.10 (  1.41) &  -51.27 (   0.58) &  -49.55 (  -1.14) &  -49.98 (  -0.71) &  -50.58 (  -0.11) \\
19&2-Methylindene (C$_{10}$H$_{10}$)                            &    33.15  &    28.61 (  4.54) &    27.31 (  5.84) &   31.79 (   1.36) &   31.38 (   1.77) &   29.09 (   4.06) &   28.53 (   4.62) \\
20&Adamantane (C$_{10}$H$_{16}$)                                &   -31.90  &   -33.90 (  2.00) &   -33.90 (  2.00) &  -31.58 (  -0.32) &  -30.87 (  -1.03) &  -30.66 (  -1.24) &  -33.07 (   1.17) \\
21&2-Methylaniline (C$_7$H$_9$N)                                &    12.72  &    12.37 (  0.35) &    11.74 (  0.98) &   13.73 (  -1.01) &   13.94 (  -1.22) &   11.42 (   1.30) &   11.26 (   1.46) \\
22&3-Methyl-2-butanone (C$_5$H$_{10}$O)                         &   -62.75  &   -62.77 (  0.02) &   -62.32 ( -0.43) &  -62.53 (  -0.22) &  -62.55 (  -0.20) &  -63.75 (   1.00) &  -63.41 (   0.66) \\
23&2-Methyl-2-butanol (C$_5$H$_{12}$O)                          &   -79.06  &   -79.32 (  0.26) &   -78.76 ( -0.30) &  -79.59 (   0.53) &  -79.29 (   0.23) &  -78.62 (  -0.44) &  -78.87 (  -0.19) \\
24&3-3-Dimethyl-2-butanone (C$_6$H$_{12}$O)                     &   -69.47  &   -70.09 (  0.62) &   -69.63 (  0.16) &  -69.34 (  -0.13) &  -69.56 (   0.09) &  -70.13 (   0.66) &  -69.92 (   0.45) \\
25&3-Hexanol (C$_6$H$_{14}$O)                                   &   -79.30  &   -80.18 (  0.88) &   -79.66 (  0.36) &  -80.85 (   1.55) &  -79.90 (   0.60) &  -80.09 (   0.79) &  -80.19 (   0.89) \\
26&2,4-Dimethylphenol (C$_8$H$_{10}$O)                          &   -38.93  &   -38.12 ( -0.81) &   -38.58 ( -0.35) &  -36.27 (  -2.66) &  -37.46 (  -1.47) &  -39.14 (   0.21) &  -39.40 (   0.47) \\
27&2-Methylpropanoic-acid (C$_4$H$_8$O$_2$)                     &  -115.70  &  -114.12 ( -1.58) &  -113.46 ( -2.24) & -114.35 (  -1.35) & -115.21 (  -0.49) & -116.19 (   0.49) & -116.21 (   0.51) \\
28&(2R,3S)-2,3-butanediol (C$_4$H$_{10}$O$_2$)                  &  -114.64  &  -111.32 ( -3.32) &  -110.53 ( -4.11) & -112.27 (  -2.37) & -112.73 (  -1.91) & -111.81 (  -2.83) & -112.36 (  -2.28) \\
29&3-Methylbutanoic acid (C$_5$H$_{10}$O$_2$)                   &  -122.45  &  -119.99 ( -2.46) &  -119.37 ( -3.08) & -120.26 (  -2.19) & -120.79 (  -1.66) & -122.19 (  -0.26) & -122.16 (  -0.29) \\
30&Hexanoic acid (C$_6$H$_{12}$O$_2$)                           &  -122.35  &  -123.13 (  0.78) &  -122.48 (  0.13) & -123.63 (   1.28) & -123.63 (   1.28) & -125.74 (   3.39) & -125.55 (   3.20) \\
31&2,2-Dimethyl-1-propanethiol (C$_5$H$_{12}$S)                 &   -30.83  &   -31.78 (  0.95) &   -32.68 (  1.85) &  -31.90 (   1.07) &  -31.83 (   1.00) &  -30.88 (   0.05) &  -30.84 (   0.01) \\
32&2-Methyl-2-pentanethiol (C$_6$H$_{14}$S)                     &   -35.44  &   -36.89 (  1.45) &   -37.60 (  2.16) &  -36.67 (   1.23) &  -36.44 (   1.00) &  -35.31 (  -0.13) &  -35.39 (  -0.05) \\
\hline
& MUE &   & 1.05 & 1.29 & 1.21  & 1.65  & 1.21  &  1.09 \\
 \hline
\end{tabular}}
    \label{tab:isomers-global}
\end{table*}

In Fig.~\ref{fig:alexbreakdown1} we inspect the transferability of HLC across various classes of compounds in G3/05 and PPE1694. 
For nonhydrogens, HLC reduces the MUE $\approx$ 
5-times for both G3/05 and the remaining nonhydrogens in PPE1694. G3/05 hydrocarbons show a 43-fold decay in MUE without the HLC correction while the decay is 27-fold for the remaining ones in PPE1694. Drop in MUE is 28- and 23-fold 
for substituted hydrocarbons in G3/05 and the rest in PPE1694, 
respectively. For inorganic hydrides, MUE drops 8-fold for G3/05, while it is 12-fold for others in PPE1694. 
Radicals show a 11-fold decay in MUE for G3/05 while the rest in PPE1694 show a 17-fold drop. Though HLC captures systematic and 
non-systematic effects successfully across compound types and datasets, we do note the quantitative prediction accuracy of G4(MP2) to drop from 1.00 kcal/mol to 1.70 kcal/mol, when going from G3/05 (Table \ref{table:small}) to PPE1694 (Table \ref{table:big}). 
This drop in accuracy when increasing the dataset size may be ascribed to both the residual uncertainties in the experimental results, and also the systematic errors introduced in the form of empirical reference data for atoms, a topic that forms the subject of Section \ref{sec:sys}.

\subsection{ISO32 dataset for isomerization reaction energies}
Accurate prediction of isomerization energies is one of the stringent tests for testing the reliability of WFTs, cWFTs and DFAs\cite{grimme2007compute}, since isomerization reaction energetics encode information about orbital hybridization, electronic conjugation, and steric effects in chemical bonding\cite{luo2011validation,sattelmeyer2006comparison}. Previous studies have observed several semi-local and hybrid DFAs to predict qualitatively incorrect trends for the energy ordering of constitutional isomers belonging for certain stoichiometries\cite{karton2012explicitly}.
%\INGREEN{"May be change the tone of the previous statement. Because although many low rung functionals perform bad but double hybrids, $\omega$functionals perform as good as MP2"}. 
From the point of view of thermochemical procedures, all systematic contributions to molecular enthalpies such as empirical atomic corrections, and HLC are cancelled in isomerization energies. Hence, prediction errors expose non-systematic errors encoded in the evaluation of electronic energies. 
To arrive at a large benchmark suite of isomerization energies, we collected constitutional isomers belonging to 32 unique stoichiometries (we denote the set ISO32).
All stoichiometries in the ISO32 dataset contains more than 7 constitutional isomers---for isomers with same $\Delta H_f^{\circ, {\rm exp.}}$, the system with lowest $\Delta H_f^{\circ, {\rm G4(MP2)}}$ was considered---amounting to 517 unique systems. 

Table \ref{tab:isomers} presents mean errors in the prediction of isomerization energies where the reactant is the global minimum per stoichiometry, amounting to 485 reaction energies. 
Due to the computational complexity, we could not perform ccCA calculations for 5 molecules: 
C$_8$H$_{14}$ (1), C$_9$H$_{16}$ (3) and C$_9$H$_{18}$ (1). The accuracies of cWFTs are found to be better than DFTs', with ccCA providing the smallest MUE of 1.86 kcal/mol and \INBLUE{$\omega$B97M-V}
showing the largest MUE, \INBLUE{2.04} kcal/mol. Interestingly, CBS-QB3 shows a similar prediction accuracy to G4;
G4(MP2) accuracy falling shortly behind. 
Benchmarking of isomerization energies benefits greatly through a joint analysis of MUE and the
Spearman rank correlation ($\rho$). 
For instance, all methods show large MUEs for the isomers of 
C$_{10}$H$_{10}$, albeit scoring a perfect $\rho=1.0$ suggesting the presence of a systematic
shift in the calculated results. In contrary, the isomers of
C$_{8}$H$_{18}$ presents a small MUE but their $\rho$ scores poorly, reaching as low as 0.53, %\INGREEN{0.53},
suggesting the isomers to be thermodynamically competitive. 
Overall, when averaging over all 485 reaction energies, we find the error trend
\INBLUE{ccCA $<$ G4 $\approx$ CBS-QB3 $<$ $\omega$B97M-D3BJ $<$ G4(MP2)  $<$ $\omega$B97M-V (see Table~\ref{tab:isomers})}
% \INGREEN{"Now the trend is: ccCA $<$ G4 $\approx$ CBS-QB3 $<$ $\omega$B97M-D3BJ $<$ G4(MP2) $<$ $\omega$B97M-V"}.
In Table~\ref{tab:isomers-global}, we present the error metrics only 
for the global minima in ISO32. 
\INBLUE{While both DFAs deliver similar average errors, one notes their 
stoichiometry-specific errors to be somewhat inconsistent.}
The general error in the modeling of global minima across stoichiometries follows 
\INBLUE{G4 $<$ $\omega$B97M-V $<$ ccCA $\approx$ $\omega$B97M-D3BJ $<$
G4(MP2) $<$ CBS-QB3}
%\INGREEN{"Now the trend is G4 $<$ $\omega$B97X-V $<$ ccCA $<$ $\omega$B97M-D3BJ $<$ G4(MP2) $<$ CBS-QB3"}
agreeing with the deep-rooted view: G4 is a better thermochemistry model than ccCA.
As stated above, ccCA's performance can be improved by extending the size of the basis sets
utilized, which however may restrict the method's applicability to large systems.

\begin{figure}[ht]
    \centering
    \includegraphics[width=8.6cm]{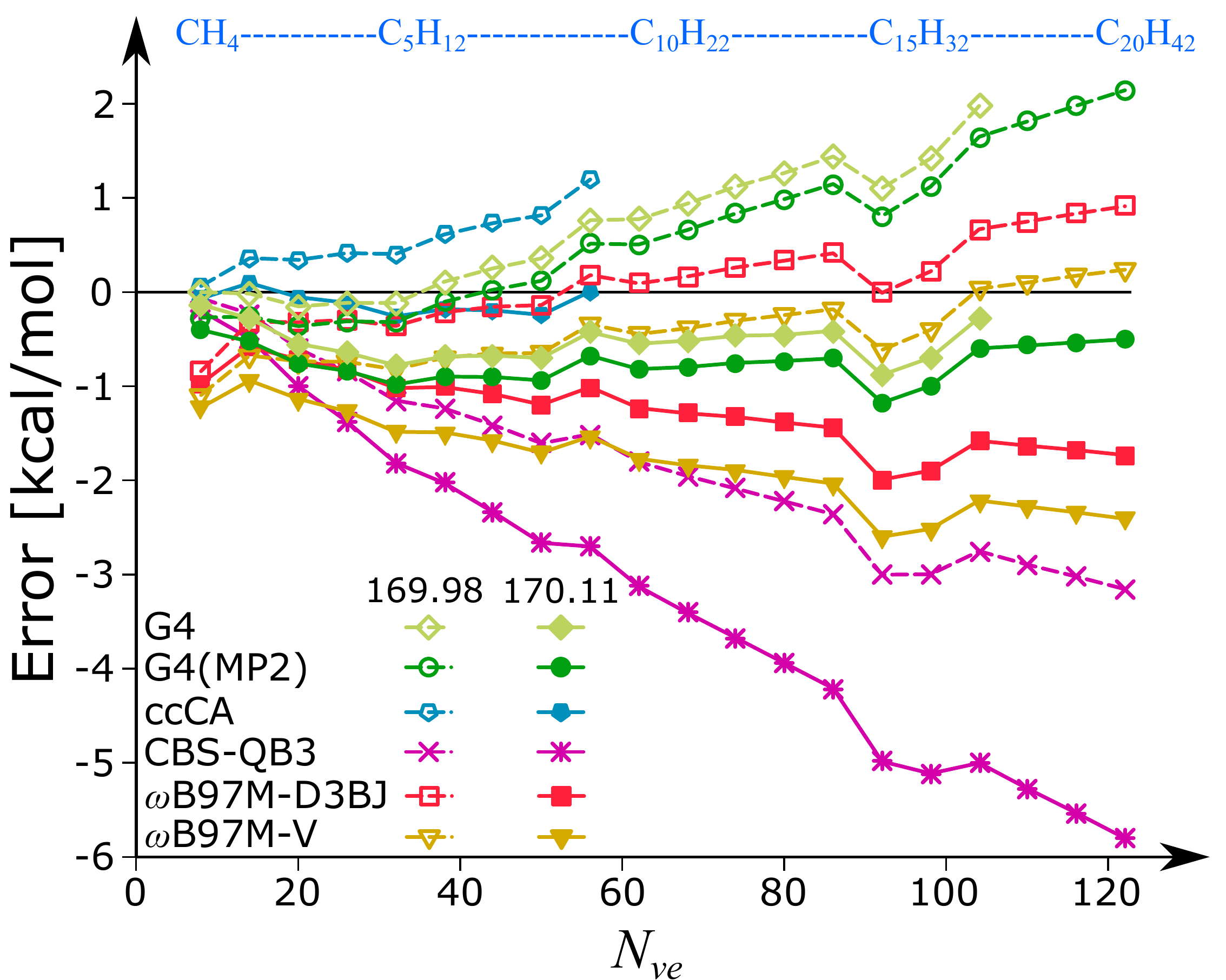}
    \caption{Systematic errors in G4, G4(MP2), ccCA, CBS-QB3, \INBLUE{$\omega$B97M-D3BJ} and \INBLUE{$\omega$B97M-V} predictions of \hof~due to the choice of $\Delta H_f^\circ ({\rm C_{gas}}, 0~{\rm K})$ shown for the smallest 20 linear alkanes.}
    \label{fig:alkanes}
\end{figure}

\subsection{Source of Systematic Errors in
Formation Enthalpies}\label{sec:sys}
Molecular $\Delta H_f^\circ (298~{\rm K})$ depends on enthalpies of constituent atoms in their 
standard elemental forms. The absolute enthalpies per atom for the elements enter the calculation as empirical constants: 
$\Delta H_f^\circ (0~{\rm K})$, which is the
zero-Kelvin enthalpy of formation; and 
the associated thermal correction, $H^\circ(298$ K$) - H^\circ(0$ K$)$. The former quantity contributes predominantly to the total enthalpy, hence any uncertainty
in its determination will be accrued with increasing molecular size. 
This has been exemplified by Tasi \textit{et al.},\cite{tasi2006origin} for C$_1$-C$_{13}$ alkanes
modelled with G2(MP2,SVP), where large systematic errors were noted when the conventional value
of $\Delta H_f^\circ (0~{\rm K})$=169.98 kcal/mol was used for the C atom. A re-evaluated value
of 170.11 kcal/mol was shown to reduce systematic errors for that dataset.

We have revisited the case of linear alkanes by increasing the set until icosane (C$_{20}$H$_{42}$) and modeling \hof~ with the four cWFTs along with \INBLUE{$\omega$B97M-D3BJ} and \INBLUE{$\omega$B97M-V} DFAs. Fig.~\ref{fig:alkanes} displays the prediction errors for all 6 methods employing two values of $\Delta H_f^\circ ({\rm C_{gas}}, 0~{\rm K})$: 169.98 kcal/mol and 170.11 kcal/mol. 
%Barring CBS-QB3, we note the deviations to drop systematically when going from the conventional parameter to the one fitted for C$_1$-C$_{13}$ alkanes in all other methods. 
We find all cWFTs except CBS-QB3 to benefit from the systematic shift resulting in reduced prediction errors---ccCA agreeing with the experimental values for C$_1$-C$_{9}$ alkanes better than G4 and G4(MP2).
The DFA, $\omega$B97M-D3BJ which showed excellent agreement with experimental values of \hof~ with an MUE of \INBLUE{0.37} kcal/mol, when using the conventional $\Delta H_f^\circ ({\rm C_{gas}}, 0~{\rm K})$ deteriorates to \INBLUE{1.29} kcal/mol upon using the re-evaluated parameter. Similarly $\omega$B97M-V's mean error rises from \INBLUE{0.48} kcal/mol 
to \INBLUE{1.82} kcal/mol when switching from $\Delta H_f^\circ ({\rm C_{gas}}, 0~{\rm K})$ = 169.98 kcal/mol to $\Delta H_f^\circ ({\rm C_{gas}}, 0~{\rm K})$ = 170.11 kcal/mol. 

% \INGREEN{$\omega$B97M-D3BJ(old): 0.37
% $\omega$B97M-D3BJ(new): 1.29, $\omega$B97M-V(old): 0.48, $\omega$B97M-V(new): 1.82}

%\INGREEN{The fall in the accuracies of DFAs $\omega$B97M-D3BJ and $\omega$B97M-V might also possible due to the atomic corrections to the predicted \hof~, fitted for $H_f^\circ ({\rm C_{gas}}, 0~{\rm K})$: 169.98 kcal/mol} 
% Seems to be a risky argument for SC.

\section{Conclusions}
A benchmark dataset of experimental \hof~ for 1,694 
compounds that are electronically and structurally rich is presented. This dataset was assembled by collecting several previously reported benchmark suites including the `legacy' dataset---G3/05 comprising 247  
entries. The resulting set included 102 
`outliers' with potentially non-negligible experimental uncertainties detected with a probabilistic approach. The procedure also takes into consideration, uncertainties associated with the reference theory, G4.

A more robust approach would be to consider more than one high-fidelity reference methods and make a joint-probabilistic model to prune the dataset. The only tunable parameter in this model is the threshold percentile used for selecting valid benchmarks; this value was set to the 95$^{\rm th}$ percentile for the reference method G4. 
We adopted a bootstrapping strategy to estimate the variance in the model arising from sampling bias. Our final results are based on the lower bound for the error-threshold to mark an entry as outlier. 

For the PPE1694 dataset, we have presented extensive benchmark results of formation enthalpies with 4 cWFTs, \textit{i.e.}, G4, G4(MP2), ccCA and CBS-QB3 as well as with 23 DFAs.  
Conventional error metrics such as MUE have been reported along with probabilistic metrics such as Q50, Q75, Q90, and Q95, that provide information about the probability and cumulative densities of errors as suggested in other studies\cite{pernot2020probabilistic,pernot2020impact}. When compared to pruned experimental values, among the methods considered in this study,  
G4 delivers the best performance with an MUE of 1.47~kcal/mol, followed by ccCA and G4(MP2) with MUEs of 1.66~kcal/mol and 1.70~kcal/mol respectively. CBS-QB3 method has MUE of 2.82~kcal/mol. 
The semi-empirical methods PM6 and PM7, as expected, result in rather accurate \hof~ 
with MUEs $\approx4$ kcal/mol. 
The most popular DFA, B3LYP exhibits a MUE of over \INBLUE{4}~kcal/mol. However, its long-range and dispersion corrected versions, namely, CAM-B3LYP and B3LYP-D3 exhibit MUEs \INBLUE{$\approx 3$} kcal/mol, while \INBLUE{ the GGA method BLYP}, mGGA method TPSS, \INBLUE{and the hybrid method X3LYP} have MUEs in the $4-7$ kcal/mol window. Dispersion corrected double hybrid functionals B2PLYP-D3 and mPW2PLYP-D show better performances with overall MUEs in \INBLUE{$2-3$} kcal/mol range. For the prediction of \hof, we found the best performing class of DFAs to be the range-separated method $\omega$B97X, and its modifications yielding rather accurate predictions with MUE $\approx 2$  kcal/mol. 
As a general trend, we note empirical dispersion corrections to improve prediction accuracy.

Our analyses revealed that the original G4(MP2) method and the empirical parameterization of the HLC involved in that model retain their transferability going from the G3/05 set with 270 entries to the proposed PPE1694 set that is 6
times larger. This suggests that the prediction accuracy of G4(MP2) to hold not only for closed-shell, organic molecules of the type encountered in the QM9 dataset, \cite{ramakrishnan2014quantum,narayanan2019accurate} but fairly well to datasets with free-radicals, non-hydrogens and inorganic hydrides. 
%Interestingly, $\omega$B97M-V, with a modest MUE of 1.92 kcal/mol in PPE1694, shows \INBLUE{non-negligible} systematic errors when tested on linear alkanes with $1-20$ C atoms. 
%This suggests that the parameters that enter into the design of this functional are somewhat overfitted to the training data with small molecules. 
It will be compelling to see if modern functionals can be designed by benchmarking on the diverse dataset presented here. Similarly, it will be of interest to see if cWFTs based on additional parameterization such as in the G4(MP2)-6X method\cite{chan2010g4} and its variants\cite{chan2019g4,semidalas2020canonical} preserve their transferability going from the small molecules set to the PPE1694 set. 

Further, from the entire benchmark suite presented here, we identified 32 sets of constitutional isomers; reaction enthalpies for this ISO32 
dataset were benchmarked over G4, G4(MP2), ccCA, CBS-QB3, and the two best performing DFAs. 
For the prediction of 485  
reaction energies, ccCA performs the best with an MUE of 1.86 kcal/mol followed by G4 and G4(MP2) with MUEs 1.94 kcal/mol and 2.00 kcal/mol respectively. 
In this case, the DFAs $\omega$B97M-D3BJ and $\omega$B97M-V have also yielded excellent predictions with MUEs $\approx 2$ kcal/mol.  
When extending the application of G4(MP2) method to molecules with large number of C atoms, we find the prediction accuracy to be sensitive to an empirical atomic parameter, which suggests that a careful evaluation of these parameters for all atom types is essential to prevent systematic accumulation of errors. Such calibration could be done with a high-fidelity method such as W4 or HEAT-456(Q). Preferably such an effort could be undertaken by pruning the total dataset and retaining only highly-precise experimental entries. Thermochemistry modeling done at such a rigor has been shown to be sensitive even to the effect anharmonicity has on molecular ZPVE\cite{pfeiffer2013anharmonic,peterson2012chemical}, hence these effects must be incorporated with methods such as second-order vibrational perturbation theory (VPT2)\cite{barone2005anharmonic,ramakrishnan2015semi}.

\section{Acknowledgements}
SKD is grateful to TIFR~Hyderabad for a junior research fellowship. This project was funded by intramural funds at TIFR Hyderabad from the Department of Atomic Energy (DAE). All calculations have been performed using the Helios computer cluster, which is an integral part of the MolDis Big Data facility, TIFR Hyderabad \href{https://moldis.tifrh.res.in/}{(https://moldis.tifrh.res.in/)}.

\section{Data Availability}
The data that support the findings of this study are openly available in the MolDis repository, \href{http://moldis.tifrh.res.in/data/prunedHOF}{http://moldis.tifrh.res.in/data/prunedHOF}. 
The same information may also be obtained from the authors through an email request.
%Additional information are available in the supplementary information.

\section*{APPENDIX: Empirical parameters}

\begin{table}[!h]
        \centering
\caption{
Heats of formation of atoms at $0$ K, $\Delta H_f^\circ (0$ K$)$, and
enthalpy corrections, for elements in their standard states, $ H^\circ(298$ K$) - H^\circ(0$ K$)$. For H, Li, Be, B, C, N, O, F, Na, Mg, Al, Si, P, S and Cl,
we used values from \RRef{curtiss1997assessment}. 
For elements
with multiple entries the most recent one, 
marked by bold font, is used. All values in kcal/mol.
}
        \begin{tabular}{ l c c }
                \hline
Atom~~~~~~~~~~& $\Delta H_f^\circ (0$ K$)$~~~~~~~~~~~~~~& $ H^\circ(298$ K$) - H^\circ(0$ K$)$\\
                \hline
                %H & 51.63 & 1.01 \\
                %Li & 37.69 & 1.10 \\
                %Be & 76.48 & 0.46 \\
                %B & 136.2 & 0.29 \\
                %C & 169.98 & 0.25 \\
                %N & 112.53 & 1.04 \\
                %O & 58.99 & 1.04 \\
                %F & 18.47 & 1.05 \\
                %Na & 25.69 & 1.54 \\
                %Mg & 34.87 & 1.19 \\
                %Al & 78.23 & 1.08 \\
                %Si & 106.6 & 0.76 \\
                %P & 75.42 & 1.28 \\
                %S & 65.66 & 1.05 \\
                %Cl & 28.59 & 1.10 \\
        K & 21.4830\cite{NIST-JANAF} & 1.6926\cite{NIST-JANAF} \\
        Ca & 42.3850\cite{NIST-JANAF} & 1.3709\cite{NIST-JANAF} \\
        Ga & 64.7633\cite{NIST-JANAF} & 1.3291\cite{NIST-JANAF} \\
                Br & 28.1836\cite{trogolo2015benchmark} & 2.930 \cite{trogolo2015benchmark,cox1989codata} \\
                Ge & \textbf{89.354}\cite{mayer1997heats}, 88.2\cite{ruscic1990photoionization}  & 1.104 \cite{wagman1982halow,mayer1997heats} \\
                As & \textbf{68.86}\cite{feller2011thermodynamic}, 68.8\cite{mayer1997heats,binning1990theoretical} &  1.23\cite{wagman1982halow} \\
                Se & 57.899\cite{wang2007gas} & 1.319\cite{wang2007gas} \\ 
        \hline
        \end{tabular}
        \label{table:TC}
\end{table}

\section{References}
\bibliography{Literature}

\end{document}